\documentclass[preprint,aps,prc,tightenlines,floatfix,showpacs,showkeys,nofootinbib]{revtex4}

\usepackage{graphicx}
\usepackage{dcolumn}
\usepackage{bm}
\usepackage{longtable}

\newcommand{\be}{\begin{equation}}
\newcommand{\ee}{\end{equation}}
\newcommand{\bea}{\begin{eqnarray}}
\newcommand{\eea}{\end{eqnarray}}

\begin{document}

\title{Nuclear matrix elements for neutrinoless double-beta decay and double-electron capture}
%
%
\author{        Amand~Faessler}
\affiliation{   Institute of Theoretical Physics,
				University of Tuebingen, 
               72076 Tuebingen, Germany}

\author{        Vadim~ Rodin}
\affiliation{   Institute of Theoretical Physics,
				University of Tuebingen, 
               72076 Tuebingen, Germany}

\author{        Fedor~\v{S}imkovic}
\affiliation{	Department of Nuclear Physics and Biophysics,
				Comenius University, 
				Mlynsk\'a dolina F1, SK--842 15 Bratislava, Slovakia}
\affiliation{	Bogoliubov Laboratory of Theoretical Physics, JINR, 
				141980 Dubna, Moscow region, Russia}

\begin{abstract}
A new generation of neutrinoless double beta decay ($0\nu\beta\beta$-decay) experiments 
with improved sensitivity is currently under design and construction. They will probe
inverted hierarchy region of the neutrino mass pattern. There is also a revived interest
to the resonant neutrinoless double-electron capture ($0\nu$ECEC), which has also a 
potential to probe lepton number conservation and to investigate the neutrino nature 
and mass scale. The primary concern are the nuclear matrix elements. Clearly, 
the accuracy of the  determination of the effective Majorana neutrino mass from the measured 
$0\nu\beta\beta$-decay half-life is mainly determined by our knowledge of the 
nuclear matrix elements. We review recent progress achieved in the calculation of $0\nu\beta\beta$ 
and $0\nu$ECEC nuclear matrix elements within the quasiparticle random phase approximation. 
A considered self-consistent approach  allow to derive the pairing, residual interactions 
and the two-nucleon short-range correlations from the same modern 
realistic nucleon-nucleon potentials. The effect of nuclear deformation is taken 
into account. A possibility to evaluate $0\nu\beta\beta$-decay matrix elements 
phenomenologically is discussed.
\end{abstract}
\medskip
\maketitle

\section{Introduction \label{SecI}}

The physics community faces a challenging problem, 
finding whether neutrinos are indeed Majorana particles (i.e. identical 
to its own antiparticle) as many particle models suggest or Dirac particles 
(i.e. is different from its antiparticle). The best sensitivity on small 
Majorana neutrino masses can be reached in the investigation of neutrinoless 
double-beta decay ($0\nu\beta\beta$-decay) \cite{AEE08,ves12},
\begin{equation}
(A,Z) \rightarrow (A,Z+2) + e^- + e^-
\label{0nbb}
\end{equation}
and the resonant neutrinoless double-electron capture ($0\nu$ECEC) \cite{mik11,ves12},
\begin{equation} 
e^-_b + e^-_b + (A,Z) \rightarrow (A,Z-2)^{**}.  
\label{0necec} 
\end{equation} 
A double asterisk in Eq.~(\ref{0necec}) means that, in general, the 
final atom $(A,Z-2)$ is excited with respect  to both the electron shell, 
due to formation of two vacancies for the electrons, and the nucleus. 
Observing the $0\nu\beta\beta$-decay and/or $0\nu$ECEC
would tell us that the total lepton number is not a conserved 
quantity and that neutrinos are massive Majorana fermions. 

The search for the $0\nu\beta\beta$-decay represents the new frontiers of 
neutrino physics, allowing in principle to fix the neutrino mass scale, 
the neutrino nature and possible CP violation effects. 
There are few tenths of nuclear systems \cite{zdes}, which offer 
an opportunity to study the $0\nu\beta\beta$-decay and the most 
favorable are those with a large $Q_{\beta\beta}$-value.

Neutrinoless double beta decay has not yet been found. 
The strongest limits on the half-life $T_{1/2}^{0\nu}$ of the 
$0\nu\beta\beta$-decay were set in Heidelberg-Moscow 
(${}^{76}\mbox{Ge}$, $1.9 \times 10^{25}$ y) \cite{bau99}, 
NEMO3 (${}^{100}\mbox{Mo}$, $1.0 \times 10^{24}$ y)  
\cite{tre11},  CUORICINO (${}^{130}\mbox{{Te}}$, $3.0 \times 10^{24}$ y)
\cite{te130} and KamLAND-Zen (${}^{136}\mbox{Xe}$, $5.7 \times 10^{24}$ y)
\cite{kamlandzen} experiments. However, there is  an unconfirmed, 
but not refuted, claim of evidence for neutrinoless double decay in $^{76}\text{Ge}$
by some participants of the Heidelberg-Moscow collaboration \cite{evidence2} 
with half-life $T^{0\nu}_{1/2} = 2.23^{+0.44}_{-0.31}\times 10^{25}$ years. 
It is expected that the GERDA experiment \cite{Gerda} in a first phase 
will check this result relatively soon.

The main aim of experiments on the search for $0\nu\beta\beta$-decay is 
the measurement of the effective Majorana neutrino mass  $m_{\beta \beta}$ 
\begin{equation}
m_{\beta\beta} = \sum_j  U^2_{ej} m_j,
\end{equation}
where $U_{ej}$ is the element of Pontecorvo-Maki-Nakagawa-Sakata (PMNS) unitary 
mixing matrix and $m_j$ is the mass of neutrino. For the most discussed case 
of mixing of three massive neutrinos (j=1,2,3) the PMNS matrix contains three 
charge parity (or CP) violating phases by assuming neutrinos to be  Majorana 
particles.

The effective Majorana neutrino mass can be calculated by using neutrino 
oscillation parameters: an assumption about the mass of lightest neutrino, 
by chosing  a 
type of spectrum (normal or inverted) and  values of CP-violating phases. 
In future experiments, CUORE (${^{130}Te}$), EXO, KamLAND-Zen (${^{136}Xe}$), 
MAJORANA (${^{76}Ge}$), SuperNEMO (${^{82}Se}$), SNO+ (${^{150}Nd}$),  
and others \cite{AEE08,ves12}, a sensitivity
\begin{equation}\label{sensitiv}
|m_{\beta\beta}|\simeq \mathrm{a~few~tens~of~meV}
\end{equation}
is planned to be reached. This is the region of the inverted hierarchy 
of neutrino masses. In the case of the normal mass hierarchy $|m_{\beta\beta}|$
is too small, a few meV, 
to be probed in the $0\nu\beta\beta$-decay 
experiments of the next generation.

We note that it is reasonable to hope that the search  for the $0\nu$ECEC 
of atoms, which are sufficiently long lived to conduct  a practical experiment, 
may also establish the Majorana nature of neutrinos. This possibility is 
considered as alternative and complementary to searches for the
$0\nu\beta\beta$-decay.

To interpret the data from the $0\nu\beta\beta$-decay and the $0\nu$ECEC  
(neutrinoless double electron capture) accurately a better understanding 
 of the nuclear structure effects important 
for the description of the nuclear matrix elements  (NMEs) is needed.
In this connection it is crucial  to develop and advance theoretical methods 
capable to  evaluate reliably NMEs, and to realistically 
assess their uncertainties.

\section{$0\nu\beta\beta$-decay NMEs: two-nucleon short-range correlations and uncertainties}

The inverse value of the $0\nu\beta\beta$-decay half-life for a given isotope $(A,Z)$ 
can be written as
\begin{eqnarray}
\frac{1}{T^{0\nu}_{1/2}} &=& \left|\frac{m_{\beta\beta}}{m_e}\right|^2~|{M'}^{0\nu}|^2~
G^{0\nu}(E_0,Z). 
\end{eqnarray}
Here, $G^{0\nu}(E_0,Z)$ and ${M'}^{0\nu}$ are, respectively,
the known phase-space factor ($E_0$ is the energy release)
and the nuclear matrix element, which depends on the nuclear 
structure of the particular isotopes $(A,Z)$, $(A,Z+1)$ and $(A,Z+2)$ under study. 
The phase space factor $G^{0\nu}(E_0,Z)$ includes fourth
power of unquenched axial-vector coupling constant $g_A$ and the inverse
square of the nuclear radius $R^{-2}$, compensated by the
factor $R$ in ${M'}^{0\nu}$. The assumed value of the 
nuclear radius is $R = r_0 A^{1/3}$ with $r_0 = 1.2~fm$.

The nuclear matrix element ${M'}^{0\nu}$ is defined as 
\begin{equation}
{M'}^{0\nu} =  \left(\frac{g^{eff}_A}{g_A}\right)^2 {M}^{0\nu}.
\label{nmep}
\end{equation}
Here, $g^{eff}_A$ is the quenched axial-vector coupling constant.
This definition of ${M'}^{0\nu}$ \cite{Rod05} allows
to display the effects of uncertainties in $g^{eff}_A$ and to use
the same phase factor $G^{0\nu}(E_0,Z)$ when calculating 
the $0\nu\beta\beta$-decay rate.

The nuclear matrix element ${M}^{0\nu}$  
consists of the Fermi (F), Gamow-Teller (GT) and tensor (T) parts as
\begin{eqnarray}
{M}^{0\nu} &=&  - \frac{M^{0\nu}_{F}}{(g^{eff}_A)^2} + M^{0\nu}_{GT} - M^{0\nu}_T \nonumber\\
&=& \langle 0^+_i|\sum_{kl} \tau^+_k \tau^+_l
[ -\frac{H_F(r_{kl})}{(g^{eff}_A)^2} + H_{GT}(r_{kl}) \sigma_{kl}
- H_T(r_{kl}) S_{kl}]
|0^+_f\rangle .\nonumber\\
\end{eqnarray}
Here
\begin{equation}
S_{kl} = 3({\vec{ \sigma}}_k\cdot \hat{{\bf r}}_{kl})
       ({\vec{\sigma}}_l \cdot \hat{{\bf r}}_{kl})
      - \sigma_{kl},\quad
\sigma_{kl}={\vec{ \sigma}}_k\cdot {\vec{ \sigma}}_l.
\label{Eq:tensor}
\end{equation}
The radial parts of the exchange potentials are
\begin{eqnarray}
H_{F,GT,T}(r_{kl}) &=& \frac{2}{\pi} R
\int_0^\infty \frac{j_{0,0,2}(q r_{kl}) h_{F,GT,T}(q^2) q}{q + \overline{E}}
dq.
\end{eqnarray}
where $R$ is the nuclear radius and $\overline{E}$ is the average energy of the virtual intermediate states
used in the closure approximation. The closure approximation  is adopted in all
the calculation of  the NMEs relevant for $0\nu\beta\beta$-decay  with the exception
of the QRPA.  The functions $h_{F,GT,T}(q^2)$ are given by \cite{anatomy}
\begin{eqnarray}
h_F(q^2) &=& f^2_{V}(q^2), \nonumber\\
h_{GT}(q^2) &=& \frac{2}{3} f^2_V(q^2) \frac{(\mu_p-\mu_n)^2}{(g^{eff}_A )^2} \frac{q^2}{4 m^2_p} + 
\nonumber\\
&& f^2_A(q^2)
\left(1 - \frac{2}{3}\frac{q^2}{q^2+m_\pi^2 } + \frac{1}{3} \frac{q^4}{(q^2+ m^2_\pi )^2}\right),
\nonumber\\
h_{T}(q^2) &=& \frac{1}{3} f^2_V(q^2) \frac{(\mu_p-\mu_n)^2}{(g^{eff}_A )^2} \frac{q^2}{4 m^2_p} + 
\nonumber\\
&&\frac{1}{3} f^2_A(q^2)\left(2 \frac{{q}^{2}}{(q^2 + m^2_\pi)}-\frac{{q}^{4}}{(q^2 + m_\pi^2)^2}\right).
\label{a14}   
\end{eqnarray}
For the vector normalized to unity and axial-vector form factors the usual dipole approximation is adopted:
$f_V({q}^{2}) = 1/{(1+{q}^{2}/{M_V^2})^2}$, $f_A({q}^{2}) = 1/{(1+{q}^{2}/{M_A^2})^2}$. 
$M_V$ = 850 MeV, and $M_A$ = 1086 MeV.  The difference in 
magnetic moments of proton and neutron is $(\mu_p-\mu_n)= 4.71$, and $g_A$ = 1.254 is assumed. 

The above definition of the ${M}^{0\nu}_\nu$ includes contribution of the higher order terms of the 
nucleon current, and   the Goldberger-Treiman PCAC 
relation, $g_P({q}^{2}) = {2 m_p g_A({q}^{2})}/({{q}^{2} + m^2_\pi})$ was employed for the induced pseudoscalar term.

\begin{table}[!t]  
  \begin{center}  
    \caption{Averaged $0\nu\beta\beta$ nuclear matrix elements 
$\langle {M'}^{0\nu} \rangle$ and their variance $\sigma$ 
(in parentheses) calculated within the QRPA and
the RQRPA. Different types of two-nucleon short-range correlations (SRC)
are considered: Milller-Spencer Jastrow SRCs (Jastrow) \cite{Rod05}; Fermi hypernetted
chain SRCc (FHCh);  unitary correlation operator method SRCs (UCOM) \cite{anatomy};
the coupled cluster method SRCs derived from the Argonne and 
CD-Bonn potentials \cite{src09}.
Three sets of single particle level schemes are used, ranging in
size from 9 to 23 orbits. 
The strength of the particle-particle interaction is adjusted 
so the experimental value of the $2\nu\beta\beta$-decay nuclear 
matrix element is correctly reproduced. 
Both free nucleon ($g^{eff}_A = g_A = 1.254$) and quenched ($g^{eff}_A = 1.0$) values 
of axial-vector coupling constant are considered. We note that unlike in
Refs. \cite{Rod05,anatomy,src09} $r_0 = 1.2$ fm instead of $r_0=1.1$ fm is assumed. 
}  
\label{tab:res}  
\begin{tabular}{lcccccccc}  
\hline\hline  
 Nucleus & $g_A^{eff}$ & meth. & 
 \multicolumn{6}{c} {$\langle {M'}^{0\nu} \rangle$} \\
 \cline{4-9}  
transition & & & 
 \multicolumn{3}{c} {SRC} & &  
 \multicolumn{2}{c} {CCM SRC} \\ 
 \cline{4-6}  \cline{8-9}  
& &  
& \hspace{0.3cm} Jastrow \hspace{0.3cm} & \hspace{0.3cm} FHCh \hspace{0.3cm}  & \hspace{0.3cm} UCOM \hspace{0.3cm}  &
& \hspace{0.3cm} Argonne \hspace{0.3cm} & \hspace{0.3cm} CD-Bonn \hspace{0.3cm}  \\  
\hline   

$^{76}Ge$
  & 1.25 &  QRPA  & 4.92(0.19) & 5.15(0.17) & 5.98(0.27) & & 6.34(0.29) & 6.89(0.35) \\
  &	 & RQRPA  & 4.28(0.13) & 4.48(0.13) & 5.17(0.20) & & 5.42(0.21) & 5.93(0.25) \\
  & 1.00 &  QRPA  & 4.18(0.15) & 4.36(0.15) & 4.97(0.23) & & 5.20(0.22) & 5.63(0.27) \\
  &	 & RQRPA  & 3.77(0.14) & 3.94(0.13) & 4.47(0.20) & & 4.59(0.15) & 5.04(0.24) \\

$^{82}Se$     
  & 1.25 &  QRPA  & 4.39(0.16) & 4.57(0.16) & 5.32(0.23) & & 5.66(0.26) & 6.16(0.29) \\
  &	 & RQRPA  & 3.81(0.14) & 3.97(0.14) & 4.59(0.17) & & 4.84(0.21) & 5.30(0.22) \\
  & 1.00 &  QRPA  & 3.59(0.13) & 3.74(0.13) & 4.29(0.19) & & 4.57(0.20) & 4.89(0.22) \\
  &	 & RQRPA  & 3.17(0.10) & 3.32(0.10) & 3.79(0.13) & & 4.00(0.15) & 4.29(0.16) \\

$^{96}Zr$     
  & 1.25 &  QRPA  & 1.22(0.03) & 1.23(0.04) & 1.77(0.02) & & 2.07(0.10) & 2.28(0.03) \\
  &	 & RQRPA  & 1.31(0.15) & 1.33(0.15) & 1.77(0.02) & & 2.01(0.17) & 2.19(0.22) \\
  & 1.00 &  QRPA  & 1.32(0.08) & 1.34(0.07) & 1.73(0.10) & & 1.90(0.12) & 2.11(0.12) \\
  &	 & RQRPA  & 1.22(0.12) & 1.24(0.12) & 1.57(0.14) & & 1.69(0.13) & 1.88(0.16) \\

$^{100}Mo$     
  & 1.25 &  QRPA  & 3.64(0.21) & 3.73(0.21) & 4.71(0.28) & & 5.18(0.36) & 5.73(0.34) \\
  &	 & RQRPA  & 3.03(0.21) & 3.12(0.21) & 3.88(0.26) & & 4.20(0.34) & 4.67(0.31) \\
  & 1.00 &  QRPA  & 2.96(0.15) & 3.02(0.15) & 3.74(0.21) & & 4.03(0.27) & 4.44(0.24) \\
  &	 & RQRPA  & 2.55(0.13) & 2.63(0.13) & 3.20(0.17) & & 3.43(0.25) & 3.75(0.21) \\

$^{116}Cd$     
  & 1.25 &  QRPA  & 2.99(0.21) & 3.11(0.21) & 3.74(0.12) & & 3.86(0.29) & 4.35(0.16) \\
  &	 & RQRPA  & 2.64(0.17) & 2.75(0.19) & 3.21(0.22) & & 3.34(0.24) & 3.72(0.26) \\
  & 1.00 &  QRPA  & 2.38(0.17) & 2.47(0.17) & 2.88(0.17) & & 2.99(0.23) & 3.31(0.21) \\
  &	 & RQRPA  & 2.14(0.14) & 2.21(0.14) & 2.55(0.17) & & 2.69(0.19) & 2.92(0.21) \\

$^{128}Te$     
  & 1.25 &  QRPA  & 3.97(0.14) & 4.15(0.15) & 5.04(0.15) & & 5.38(0.17) & 5.99(0.17) \\
  &	 & RQRPA  & 3.52(0.13) & 3.68(0.14) & 4.45(0.15) & & 4.71(0.17) & 5.26(0.16) \\
  & 1.00 &  QRPA  & 3.11(0.09) & 3.23(0.10) & 3.88(0.11) & & 4.11(0.13) & 4.54(0.13) \\
  &	 & RQRPA  & 2.77(0.09) & 2.88(0.09) & 3.44(0.10) & & 3.62(0.12) & 4.00(0.12) \\

$^{130}Te$  
  & 1.25 &  QRPA  & 3.56(0.13) & 3.72(0.14) & 4.53(0.12) & & 4.77(0.15) & 5.37(0.13) \\
  &	 & RQRPA  & 3.22(0.13) & 3.36(0.15) & 4.07(0.13) & & 4.27(0.15) & 4.80(0.14) \\
  & 1.00 &  QRPA  & 2.55(0.08) & 2.93(0.08) & 3.52(0.07) & & 3.69(0.11) & 4.11(0.08) \\
  &	 & RQRPA  & 2.15(0.14) & 2.66(0.09) & 3.17(0.08) & & 3.29(0.11) & 3.69(0.09) \\

$^{136}Xe$ 
  & 1.25 &  QRPA  & 2.16(0.13) & 2.25(0.12) & 2.73(0.13) & & 2.88(0.14) & 3.23(0.14) \\
  &	 & RQRPA  & 2.02(0.12) & 2.11(0.14) & 2.54(0.15) & & 2.68(0.16) & 3.00(0.17) \\
  & 1.00 &  QRPA  & 1.70(0.09) & 1.77(0.09) & 2.12(0.11) & & 2.21(0.10) & 2.47(0.09) \\
  &	 & RQRPA  & 1.59(0.09) & 1.66(0.10) & 1.97(0.11) & & 2.06(0.11) & 2.30(0.12) \\

\hline\hline  
\end{tabular}  
  \end{center}  
\end{table}  

The nuclear matrix elements ${M'}^{0\nu}$ for the $0\nu\beta\beta$ decay must be 
evaluated using tools of nuclear structure theory. Unfortunately, there are no 
observables that could be simply and directly linked to the magnitude of 
$0\nu\beta\beta$ nuclear matrix elements and that could be used to determine 
them in an essentially model independent way.

During many years two  approaches were used: the Quasiparticle Random
Phase Approximation (QRPA)\cite{Rod05,anatomy,suh_UCOM} and the Interacting
Shell Model (ISM)\cite{men09}. There are substantial
differences between both approaches. The QRPA treats a large single
particle model space, but truncates heavily the included configurations.
The ISM, by contrast, treats a small fraction of this model space,
but allows the nucleons to correlate in many different ways.
In the last few years several new approaches have been used for
the calculation of the $0\nu\beta\beta$-decay NMEs: the angular
momentum Projected Hartree-Fock-Bogoliubov method (PHFB)
\cite{Hir08}, the Interacting Boson Model (IBM) \cite{Bar09},
and the Energy Density Functional method (EDF) \cite{Rodri10}.

The standard QRPA method consists of two steps. First, the mean field 
corresponding to the  minimum of energy is determined and the 
like-particle pairing interaction is taken into account by employing 
the quasiparticle representation. In the second step the linearized
equations of motion are solved in order to describe small amplitude 
vibrational-like modes around that minimum. In the renormalized version 
of QRPA (RQRPA) the violation of the Pauli exclusion principle is partially corrected. 

The drawback of QRPA is the fact that, unlike in BCS, the particle number 
is not conserved automatically, even on average. For realistic Hamiltonians 
the differences between averaged particle numbers on the RPA ground state
and the exact particle numbers could be of the order of unity 
(an extra or missing neutron or proton). The selfconsitent renormalized QRPA 
method (SRQRPA) removes this drawback by treating the BCS and QRPA vacua 
simultaneously \cite{occup}. For the neutron-proton systems the method 
was proposed and tested on the exactly solvable simplified models in Ref. 
\cite{Delion}. It is a generalization of the
procedure proposed earlier in \cite{Schuck}.

In the QRPAthe phonon operators are defined as
\begin{equation}
Q^{\dagger (k)}_{J,M} = \Sigma_{pn} [ X_{(pn) J}^k A^{\dagger}_{(pn) J,M} -
Y_{(pn) J}^k \tilde{A}_{(pn) J,M} ] ~,
\label{eq:phonon}
\end{equation}
where $X_{(pn) J}^k$ and $Y_{(pn) J}^k$ are the usual variational amplitudes,
and $A^{\dagger}_{(pn) J,M}$ is the angular momentum coupled two-quasiparticle 
creation operator. $p,n$ signify the quantum numbers of the proton, respectively neutron, orbits.
The $X$ and $Y$ amplitudes, as well as  the corresponding
energy eigenvalues $\omega_k$ are determined by solving the QRPA eigenvalue equations
for each $J^{\pi}$
\begin{equation}
\left( \begin{array}{cc}
A & B \\ -B & - A 
\end{array} \right)
\left( \begin{array}{c}
X \\ Y \end{array} \right)
= \omega \left(
 \begin{array}{c}
X \\ Y \end{array} \right) ~.
\label{eq_rpa}
\end{equation}
The matrices $A$ and $B$ above are determined by the Hamiltonian rewritten in terms of
the quasiparticle operators:
\begin{eqnarray}
A^J_{pn,p'n'}  & = & \langle O | (a^{\dagger}_p a^{\dagger}_n )^{(JM) ^{\dagger}} \hat{H}
(a^{\dagger}_{p'} a^{\dagger}_{n'} )^{(JM)} | O \rangle \\
\nonumber
B^J_{pn,p'n'}  & = & \langle O |  \hat{H} (a^{\dagger}_p a^{\dagger}_n )^{(J - M)} (-1)^M 
(a^{\dagger}_{p'} a^{\dagger}_{n'} )^{(JM)} | O \rangle
\label{eq_ABdef}
\end{eqnarray}
Here, $| O \rangle$ is the BCS vacuum state. 

In the RQRPA and SRQRPA instead of $| O \rangle$  the correlated QRPA
ground state $| 0^+_{QRPA} \rangle$ is considered. Then 
instead of the standard $X$ and $Y$
everywhere and also in the QRPA equations 
of motion the renormalized amplitudes are used:
\begin{equation}
{\overline{X}}^m_{(pn, J^\pi)}  =  {\cal D}^{1/2}_{pn}~ 
X^m_{(pn, J^\pi)}, ~~~~~
{\overline{Y}}^m_{(pn, J^\pi)}  =  {\cal D}^{1/2}_{pn}~ 
Y^m_{(pn, J^\pi)}, 
\end{equation}
where  renormalization factors ${\cal D}_{pn}$ are given by
\begin{eqnarray}
\label{eq:QRPA}
{\cal D}_{pn}
 &  = &  
\langle  0^+_{QRPA}|[A^{}_{(pn) J,M}, A^{\dagger}_{(pn) J,M}]| 0^+_{QRPA} \rangle
= 1 -  \xi_p - \xi_n  \\  \nonumber
&= & 1 -  \frac{1}{2j_p + 1} 
\Sigma_{n'} {\cal D}_{pn'} 
\left( \Sigma_{J,k} (2J+1) | \overline{Y}_{pn'}^{J,k} |^2 \right)\nonumber
\\  \nonumber
&&~~ -  \frac{1}{2j_n + 1} \Sigma_{p'} {\cal D}_{p'n} 
\left( \Sigma_{J,k} (2J+1) | \overline{Y}_{p'n}^{J,k} |^2 \right).
\end{eqnarray}
Here, $\xi_{n(p)}$ is the expectation value of the number of quasiparticles 
in the orbit $n(p)$, 
\begin{eqnarray}
\xi_{n(p)}
\equiv  \frac{\langle 0^+_{QRPA} | \left[a^+_{n(p)} a_{n(p)}\right]_{00} | 0^+_{QRPA} \rangle}
{\sqrt{2 j_{n(p)} +1 }}.
\end{eqnarray}
$a^+_{j_{n(p)},m}, a_{j_{n(p)},m}$
are the creation and annihilitation operators for the quasiparticle 
with quantum numbers $n(p),m$. The renormalizattion coefficients ${\cal D}_{pn}$
and the quasiparticle occupation numbers $\xi_j$ can be obtained iteratively 
using the equations of motion of the (S)RQRPA. 

In the correlated QRPA ground state the occupation numbers are no longer the 
pure BCS quantities. Instead, they depend, in addition, on the solutions of the
QRPA equations of motion for all multipoles $J$, and can be evaluated using
\begin{eqnarray}
{\rm n}^{QRPA}_ {n(p)}& = & \langle 0^+_{QRPA} |    \Sigma_m c^+_{n(p),m} c_{n(p),m} | 0^+_{QRPA} \rangle
\\ \nonumber
& \simeq & (2j_{n(p)}+ 1)~\left[ v^2_{n(p)}+ (u^2_{n(p)} - v^2_{n(p)})~ \xi_{n(p)} \right].
\end{eqnarray}
Here, $c^+_{j,m}$ is the creation operator for a proton in the orbit $j_p$ or a neutron
in the orbit $j_n$ and $c_{j,m}$ is the corresponding annihilation operator. The amplitudes 
$v_{j_p}$ and $v_{j_n}$ are obtained by solving the gap equations. 

In SRQRPA the BCS equations are reformulated. 
This is achieved by recalculating the $u$ and $v$ amplitudes from the  minimum 
condition of the RQRPA ground-state energy.
In SRQRPA thus the state around which the vibrational modes
occur is no longer the quasiparticle vacuum, but instead the
Bogoliubov transformation is chosen is such a way that provides the optimal and consistent
basis while preserving the form of the phonon operator, Eq. (\ref{eq:phonon}).

In practice, the SRQRPA equations are solved double iteratively. One begins with the standard BCS 
$u,v$ amplitudes, solves the RQRPA equations of motion and calculates the factors $D_{pn}$.
The $u,v$ amplitudes are recalculated and the procedure is repeated until the selfconsistency
is achieved. Numerically, the double iteration procedure represents a challenging problem. 
It was resolved in \cite{Benes} where instead of the G-matrix  based interaction 
the pairing part (and only that part) of the problem was replaced by a pairing interaction 
that uses a constant matrix element whose value was adjusted to reproduce the experimental 
odd-even mass differences.

In the QRPA, RQRPA, and SRQRPA the ${M}^{0\nu}$ is written as the sum
over the virtual
intermediate states, labeled by their angular momentum and parity
$J^{\pi}$ and indices $k_i$ and $k_f$:
\begin{eqnarray}
M_K  =  \sum_{J^{\pi},k_i,k_f,\mathcal{J}} \sum_{pnp'n'}
(-1)^{j_n + j_{p'} + J + {\mathcal J}} 
\sqrt{ 2 {\mathcal J} + 1}
\left\{
\begin{array}{c c c}
j_p & j_n & J  \\
 j_{n'} & j_{p'} & {\mathcal J}
\end{array}
\right\}  \times~~~~~~~~~~~~~~~~~~~
\nonumber \\
\langle p(1), p'(2); {\mathcal J} \parallel \bar{f}(r_{12})
O_K \bar{f}(r_{12}) \parallel n(1), n'(2); {\mathcal J} \rangle \times~~
\nonumber \\
\langle 0_f^+ ||
[ \widetilde{c_{p'}^+ \tilde{c}_{n'}}]_J || J^{\pi} k_f \rangle
\langle  J^{\pi} k_f |  J^{\pi} k_i \rangle
 \langle  J^{\pi} k_i || [c_p^+ \tilde{c}_n]_J || 0_i^+ \rangle ~.
\nonumber\\
\label{eq:long}
\end{eqnarray}
The operators $O_K, K$ = Fermi (F), Gamow-Teller (GT), and Tensor
(T), contain neutrino potentials and spin and isospin operators, and
RPA energies $E^{k_i,k_f}_{J^\pi}$.
Two separate multipole decompositions are built into Eq.\
(\ref{eq:long}). One is in terms the
$J^{\pi}$ of the virtual states in the intermediate nucleus, the
good quantum numbers of the QRPA and RQRPA. The other decomposition
is based on the angular momenta and parities ${\mathcal J}^{\pi}$ of the
pairs of neutrons that are transformed into protons with the same
${\mathcal J}^{\pi}$. The nucleon orbits are labeled in Eq.(\ref{eq:long}) 
by $p,p',n,n'$. 

The QRPA-like approaches do not allow to introduce short-range 
correlations (SRCs) into the two-nucleon relative wave function. The 
traditional way is to introduce an explicit Jastrow-type correlation 
function $f(r_{12})$ into the involved two-body transition 
matrix elements (see Eq. \ref{eq:long}). In the parametrization 
of Miller and Spencer \cite{Spencer} we have
\begin{equation}
f(r_{12}) = 1 - c e^{-a r^2}(1-b r^2), ~a=1.1~fm^{-2},~ b=0.68~fm^{-2}.
\end{equation}
These two parameters ($a$ and $b$) 
are correlated and chosen in the way that the norm 
of the relative wave function $|{\overline{\Psi}}_{\mathcal J}  \rangle$ 
is conserved. 

Recently, it was proposed \cite{suh_UCOM} to adopt instead of the Jastrow method 
the UCOM approach for description of the two-body correlated wave function 
\cite{UCOM}. The UCOM method produces good results for 
the binding energies of nuclei already at the Hartree-Fock 
level \cite{roth}. 

A self-consistent calculation of the $0\nu\beta\beta$-decay
NMEs in the QRPA-like approaches was performed in \cite{src09}.
The pairing and residual interactions as well as the two-nucleon 
short-range correlations were for the first time derived  from the same
modern realistic nucleon-nucleon potentials, namely from the 
charge-dependent Bonn potential (CD-Bonn) and the Argonne V18
potential. A method of choice was the coupled cluster
method (CCM) \cite{muether}. For purpose of numerical calculation 
of the $0\nu\beta\beta$-decay  NMEs the CCM short-range correlation functions 
were presented in an analytic form of Jastrow-like function as \cite{src09}
\begin{equation}
f_{A,B}(r_{12}) = 1 ~-~ c~ e^{-a r^2}(1-b r^2). 
\end{equation}
The set of parameters for Argonne  and CD-Bonn NN interactions 
is given by
\begin{eqnarray}
f_{A}(r_{12}):~~a &=& 1.59~fm^{-2},~~b = 1.45~fm^{-2},~~c = 0.92,\nonumber\\
f_{B}(r_{12}):~~a &=& 1.52~fm^{-2},~~b = 1.88~fm^{-2},~~c = 0.46.\nonumber\\
\end{eqnarray}
The calculated NMEs with these short-range correlation functions agree within 
a few percentages with those obtained without this approximation.
We note that the dependence of the SRC on the value of oscillator length $b$ 
is rather weak.  

%
%

\begin{table}[!t]
  \begin{center}
    \caption{\label{tab:t12}
The calculated ranges of the nuclear matrix element
$M^{'0\nu}$ evaluated within the QRPA (column 2), 
RQRPA (column 4) and SRQRPA (column 6), with standard ($g^{eff}_A= g_A = 1.254$) and
quenched ($g^{eff}_A = 1.0$) axial-vector couplings and with  the coupled cluster 
method (CCM) CD-Bonn and Argonne short-range correlation (SRC) functions. 
Columns 3, 5 and 7 give the  $0\nu\beta\beta$-decay 
half-life ranges corresponding to values of the matrix-elements 
in columns 2, 4 and 6 for $|m_{\beta\beta}| = 50$~meV.
$T^{0\nu -exp}_{1/2}$ is the experimental lower bound on the 
$0\nu\beta\beta$-decay half-life for a given isotope. 
}
\begin{tabular}{lccccccccc}
\hline\hline
 Nucl. & \multicolumn{2}{c}{QRPA} & &
           \multicolumn{2}{c}{RQRPA} & &
           \multicolumn{2}{c}{SRQRPA}\\ 
           \cline{2-3} \cline{5-6} \cline{8-9}
 & $M^{0\nu}$ & $T^{0\nu}_{1/2}$ [y] &
 & $M^{'0\nu}$ & $T^{0\nu}_{1/2}$ [y] &
 & $M^{'0\nu}$ & $T^{0\nu}_{1/2}$ [y] & $T^{0\nu -exp}_{1/2}$[y]  \\\hline
$^{76}Ge$
  &  $(5.0,7.2)$ & $(3.0,6.3)\times 10^{26}$ &
  &  $(4.5,6.2)$ & $(4.1,7.9)\times 10^{26}$ &
  &  $(4.3,6.2)$ & $(4.0,8.6)\times 10^{26}$ 
  &  $1.9\times 10^{25}$ \cite{bau99} \\
$^{82}Se$  
  &  $(4.4,6.4)$ & $(8.5,18.)\times 10^{25}$ &
  &  $(3.8,5.6)$ & $(1.2,2.4)\times 10^{26}$ &
  &  $(3.9,6.1)$ & $(9.5,22.)\times 10^{25}$ 
  &  $3.2\times 10^{23}$ \cite{tre11} \\
$^{100}Mo$
  &  $(3.7,6.1)$ & $(5.9,15.)\times 10^{25}$ &
  &  $(3.2,5.0)$ & $(8.8,21.)\times 10^{25}$ &
  &  $(4.0,5.5)$ & $(7.3,13.)\times 10^{25}$ 
  &  $1.0\times 10^{24}$ \cite{tre11}\\
$^{130}Te$
  &  $(3.6,5.5)$ & $(7.4,18.)\times 10^{25}$ &
  &  $(3.2,4.7)$ & $(1.0,2.2)\times 10^{26}$ &
  &  $(3.6,5.1)$ & $(8.5,17.)\times 10^{25}$ 
  &  $3.0\times 10^{24}$ \cite{te130}\\
$^{136}Xe$
  &  $(2.1,3.4)$ & $(1.9,4.8)\times 10^{26}$ &
  &  $(2.0,3.2)$ & $(2.1,5.5)\times 10^{26}$ &
  &  $(2.4,3.6)$ & $(1.6,3.7)\times 10^{26}$ 
  &  $5.7\times 10^{24}$ \cite{kamlandzen}\\
\hline\hline
\end{tabular}
  \end{center}
\end{table}

In Table \ref{tab:res} the QRPA and RQRPA results are presented separately 
for different types of two-nucleon short-range correlations (SRC)
are considered: Milller-Spencer Jastrow SRCs (Jastrow) \cite{Rod05}; Fermi hypernetted
chain SRCc (FHCh);  unitary correlation operator method SRCs (UCOM) \cite{anatomy};
the coupled cluster method SRCs derived from the Argonne and 
CD-Bonn potentials \cite{src09} based on an extension of the Brueckner theory (Coupled Cluste Method = CCM). Two different values of the axial coupling constant, 
free nucleon $g^{eff}_A=g_A=1.254$ and quenched $g^{eff}_A=1.0$, are taken into
account. The strength of the particle-particle interaction is adjusted 
so the experimental value of the $2\nu\beta\beta$-decay nuclear 
matrix element is correctly reproduced \cite{Rod05}. 
The NME calculated within this procedure, which 
includes three different model spaces, is denoted as the averaged 
$0\nu\beta\beta$-decay NME $\langle {M'}^{0\nu} \rangle$. We note that
the values of NMEs become essentially independent on the size of 
the single-particle basis and rather stable with respect to the 
possible quenching  of the $g_A$. 

From Table \ref{tab:res} it follows that the QRPA values are about
10-15\% larger in comparison with the RQRPA values. The largest NMEs are
those calculated with the CCM CD-Bonn correlation function.
In comparison with them the NMEs obtained with the CCM CD-Argonne 
correlation function and the UCOM SRCs are about 10\% smaller.
This is explained by the fact that the  CCM Argonne correlation 
function cuts out more the small $r_{12}$ part from the relative
wave function of the two-nucleons as the CCM CD-Bonn  
correlation function. The smallest in magnitude are 
matrix elements for the $0\nu\beta\beta$ decay obtained
with the traditional approach of using the Miller-Spencer Jastrow 
SRC and the Fermi hypernetted chain SRCc. 

In Table \ref{tab:t12} we show the calculated ranges of 
the nuclear matrix element $M^{'0\nu}$ evaluated within
the QRPA, RQRPA \cite{src09} and SRQRPA \cite{petcov11} 
in a self-consistent way with the CCM CD-Bonn and Argonne 
SRC functions by assuming both the standard ($g_A = 1.254$) and
quenched ($g_A = 1.0$) axial-vector couplings.
These ranges quantify the uncertainty in the calculated
$0\nu\beta\beta$-decay NMEs of a given QRPA-like approach.
By comparing the SRQRPA with the RQRPA results we conclude 
that the requirement of conserving the particle number have 
not caused substantial changes in the value of the $0\nu\beta\beta$ 
matrix elements in that case.

Given the interest in the subject, in Table \ref{tab:t12} we show 
also the range of predicted $0\nu\beta\beta$-decay half-lives 
of ${^{76}Ge}$, ${^{82}Se}$, ${^{100}Mo}$,  ${^{130}Te}$ and ${^{136}Xe}$
corresponding to full range of $M'^{0\nu}$ for $|m_{\beta\beta}|$ = 50 meV. 
This is a rather conservative range within the considered QRPA framework. 
It represents roughly a required sensitivity of the $0\nu\beta\beta$-decay 
experiment in the case of inverted hierarchy of neutrino masses, which can be 
compared with the current bound on the $0\nu\beta\beta$-decay
half-life $T^{0\nu -exp}_{1/2}$.  

\section{$\beta\beta$ decay of deformed nuclei within QRPA}


One of the best candidates for searching $0\nu\beta\beta$ decay is $^{150}$Nd since 
it has the second highest endpoint, $Q_{\beta\beta}=$3.37 MeV, and the largest phase space factor for the decay (about 33 times larger than that for $^{76}$Ge, see e.g.~\cite{Vog92}). 
The SNO+ experiment at the Sudbury Neutrino Observatory will use a Nd-loaded scintillator to search for neutrinoless double beta decay by looking for a distortion in the energy spectrum of decays at the endpoint~\cite{SNO+}.

However, $^{150}$Nd is well known to be a rather strongly deformed nucleus. This 
strongly hinders a reliable theoretical evaluation of the corresponding $0\nu\beta\beta$-decay NMEs (for instance, it does not seem feasible in the near future to reliably treat this nucleus within the large-scale nuclear shell model (ISM), see, e.g., Ref.~\cite{men09}). 
Recently, more phenomenological approaches like the pseudo-SU(3) model~\cite{Hir95}, the 
PHFB approach~\cite{Hir08}, the 
IBM~\cite{Bar09}, and the 
EDF~\cite{Rodri10} have been employed to calculate 
$M^{0\nu}$ for strongly deformed heavy nuclei (a comparative analysis of different approximations involved in some of the models can be found in Ref.~\cite{Esc10}).
The results of these models generally reveal a substantial suppression of $M^{0\nu}$ for $^{150}$Nd as compared with the QRPA result of  Ref.~\cite{Rod05} where $^{150}$Nd and $^{150}$Sm were treated as spherical nuclei. However, the calculated NMEs $M^{0\nu}$ for $^{150}$Nd show a rather significant spread.

One of the most up-to-date microscopic ways to describe the effect of nuclear deformation on $\beta\beta$-decay NMEs $M^{2\nu}$ and $M^{0\nu}$ is provided by the QRPA. A QRPA approach for calculating $\beta\beta$-decay amplitudes 
in deformed nuclei has been developed in a series of papers~\cite{Sim03,Sim04,Sal09,Fang10,Fang11}. $M^{2\nu}$ were calculated in Refs.~\cite{Sim03,Sim04} with schematic separable forces, and in Ref.~\cite{Sal09} - with realistic residual interaction. It was demonstrated in Refs.~\cite{Sim03,Sim04,Sal09} that 
deformation introduces a mechanism of suppression of the $M^{2\nu}$ matrix element 
which gets stronger when deformations of the initial and final nuclei differ from each other. A similar dependence of the suppression of both $M^{2\nu}$ and $M^{0\nu}$ matrix elements on the difference in deformations has been found in the PHFB~\cite{Hir08} and the ISM~\cite{men09}. 

In Refs.~\cite{Fang10,Fang11}, the first QRPA calculations of $M^{0\nu}$ 
with an account for nuclear deformation were done. 
The calculations showed a suppression of $M^{0\nu}$ for $^{150}$Nd by 
about 40\% as compared with our previous spherical QRPA result~\cite{Rod05}.
In the next section we review the results of Refs.~\cite{Sim03,Sim04,Sal09,Fang10,Fang11}.

\subsection{Formalism}

The NMEs $M^{2\nu}$ and $M^{0\nu}$, as the scalar measures of the decay rates, can be calculated in any coordinate system.
For strongly deformed, axially symmetric, nuclei the most convenient choice is 
the intrinsic coordinate system associated with the rotating nucleus. This employs the adiabatic Bohr-Mottelson approximation that is well justified for $^{150}$Nd, $^{160}$Gd and $^{160}$Dy, which indeed reveal strong deformations. As for $^{150}$Sm, the enhanced quadrupole moment of this nucleus is an indication for its static deformation.

Though it is difficult to evaluate the effects beyond the adiabatic approximation employed here, one might anticipate already without calculations that the smaller the deformation is, the smaller should be the deviation of the calculated observables from the ones obtained in the spherical limit.
In this connection it is worth noting that spherical QRPA results can exactly be reproduced in the present QRPA calculation by letting the deformation vanish, in spite of the formal inapplicability of the adiabatic ansatz for the wave function in this limit.

We give here for completeness the formalism of the QRPA calculations of NMEs $M^{2\nu}$ and $M^{0\nu}$ in deformed nuclei as developed in 
Refs.~\cite{Sim03,Sim04,Sal09,Fang10,Fang11}. 

Nuclear excitations in the intrinsic system $| K^\pi\rangle$ are characterized by the projection of the total angular momentum onto the nuclear symmetry axis $K$ (the only projection which is conserved in strongly deformed nuclei) and the parity $\pi$. 

The intrinsic states $| K^\pi,m\rangle$ are generated within the
QRPA by a phonon creation operator acting on the ground-state wave function: 
\begin{equation}
| K^\pi,m\rangle=Q_{m,K}^\dagger |0^+_{g.s.}\rangle;~~~~Q_{m,K}^\dagger = \sum_{pn} X^{m}_{pn, K} A^\dagger _{pn, K} - Y^{m}_{pn, K} \bar{A}_{pn,K}.
\label{3}
\end{equation}
Here, $A^\dagger _{pn,K}=a^\dagger_{p}{a}^{\dagger}_{\bar{n}}$ and $\bar{A}_{pn,K}={a}_{\bar{p}}{a}_{n}$ 
are the two-quasiparticle creation and annihilation operators, 
respectively, with the bar denoting the time-reversal operation.
The quasiparticle pairs $p\bar n$ are defined by the selection rules $\Omega_p -\Omega_n = K$ 
and $\pi_p\pi_n=\pi$, where $\pi_\tau$ is the single-particle (s.p.) parity and  $\Omega_\tau$ is the projection 
of the total s.p. angular momentum on the nuclear symmetry axis ($\tau = p,n$).
The s.p. states $|p\rangle$ and $|n\rangle$ 
of protons and neutrons are calculated by solving the Schr\"odinger equation with the deformed axially symmetric Woods-Saxon potential~\cite{Sal09}. 
In the cylindrical coordinates the  
deformed Woods-Saxon s.p. wave functions $|\tau \Omega_\tau\rangle$ with $\Omega_\tau>0$ are decomposed over the deformed harmonic oscillator 
s.p. wave functions (with the principal quantum numbers $(N n_{z} \Lambda)$) and the spin wave functions 
$|\Sigma=\pm \frac12\rangle$:
 \begin{eqnarray}
|\tau \Omega_\tau\rangle &=& \sum_{N n_z \Sigma} b_{N n_z \Sigma} |N n_z \Lambda_\tau=\Omega_\tau-\Sigma\rangle |\Sigma\rangle,
\label{tau}
\end{eqnarray}
where $N=n_\perp+n_z$ ($n_\perp=2n_\rho+|\Lambda|$), $n_z$ and $n_\rho$ are the number of nodes
of the basis functions in the $z$- and $\rho$-directions,
respectively; $\Lambda =  \Omega - \Sigma$ and $\Sigma$ are the
projections of the orbital and spin angular momentum onto the symmetry axis $z$.
For the s.p. states with the negative projection $\Omega_\tau=-|\Omega_\tau|$, 
which are degenerate in energy with $\Omega_\tau=|\Omega_\tau|$,
the time-reversed version of Eq.~(\ref{tau}) is used as a definition (see also Ref.~\cite{Sal09}).
The states $(\tau,\bar \tau)$ comprise the whole single-particle model space.

The deformed harmonic oscillator wave functions $|N n_z \Lambda\rangle$  
can be further decomposed over the spherical harmonic oscillator ones   
$|n_rl\Lambda\rangle$ by calculating the corresponding spatial  
overlap integrals $A^{n_rl}_{N n_z \Lambda} =\langle n_r l \Lambda|N n_{z} \Lambda\rangle$  
($n_r$ is the radial quantum number,  $l$ and $\Lambda$ are the orbital angular momentum and its projection onto $z$-axes, respectively), see Appendix of Ref.~\cite{Sal09} for more details.   
Thereby, the wave function (\ref{tau}) can be reexpressed as   
\begin{eqnarray} 
|\tau \Omega_\tau\rangle&=& \sum_{\eta}B^{\tau}_{\eta}|\eta\Omega_\tau\rangle , 
\label{4} 
\end{eqnarray} 
where $|\eta\Omega_\tau\rangle=\sum\limits_{\Sigma} C^{j\Omega_\tau}_{l~ \Omega_\tau-\Sigma~\frac12~\Sigma} 
|n_r l \Lambda=\Omega_\tau-\Sigma\rangle |\Sigma\rangle$ is the spherical harmonic oscillator wave function in the $j$-coupled scheme ($\eta=(n_rlj)$), and  
$B^{\tau}_{\eta}= \sum\limits_{\Sigma}C^{j\Omega_{\tau}}_{l ~\Omega_{\tau}-\Sigma~\frac12~\Sigma}\, A^{n_r l}_{N n_z \Omega_{\tau}-\Sigma}\, b_{N n_z \Sigma}$, with $C^{j\Omega_{\tau}}_{l ~\Omega_{\tau}-\Sigma~\frac12~\Sigma}$ being the Clebsch-Gordan coefficient. 

The  QRPA equations:
\begin{eqnarray}
\left( \matrix{ {\cal A}(K) & {\cal B}(K) \cr
-{\cal B}(K) & -{\cal A}(K) }\right) 
~\left( \matrix{ X^m_K \cr Y^m_K} \right)~ = ~
\omega_{K,m}
~\left( \matrix{ X^m_K \cr Y^m_K} \right),
\label{5}
\end{eqnarray}
with realistic residual interaction are solved to get the forward $X^m_{i K}$,  
backward $Y^m_{i K}$ amplitudes and the excitation energies $\omega^{m_i}_K $ and $ \omega^{m_f}_K$  of  the $m$-th $K^\pi$ 
state in the intermediate nucleus. The matrix $\cal A$ and $\cal B$ are defined by
\begin{eqnarray}
{{\cal A}_{p n,{p'}{n'}}}(K)&=&{\delta}_{p n,{p'}{n'}}(E_p+E_n)+g_{pp}(u_p u_n u_{p'} u_{n'}+ v_p v_n v_{p'} v_{n'}) 
V_{p \bar{n}p'\bar{n'}}\nonumber\\
&&~~~~~~~~~~~~~~~~~~~~~~-g_{ph}(u_p v_n u_{p'} v_{n'}+ v_p u_n v_{p'} u_{n'})
V_{p n'p'n}\nonumber\\
{{\cal B}_{p n,{p'}{n'}}}(K)&=&-g_{pp}(u_p u_n v_{p'} v_{n'}+ v_p v_n u_{p'} u_{n'}) 
V_{p \bar{n}p'\bar{n'}}\nonumber\\
&&-g_{ph}(u_p v_n v_{p'} v_{n'}+ v_p u_n u_{p'} v_{n'})
V_{p n'p'n}
\label{6}
\end{eqnarray}  
where $E_p+E_n$ are the two-quasiparticle excitation energies, 
$V_{p n,{p'}{n'}}$ and $V_{p \bar {n},{p'}\bar {n'}}$ are the $p-h$ and $p-p$
matrix elements of the residual nucleon-nucleon interaction $V$, respectively,
$u_\tau$ and $v_\tau$ are the coefficients of the Bogoliubov transformation.

As a residual two-body interaction we use the nuclear Brueckner $G$~matrix, which is a solution of the Bethe-Goldstone equation, derived from the charge-depending Bonn (Bonn-CD) one boson exchange potential, as used also in the spherical calculations of Ref.~\cite{Rod05}. 
The $G$~matrix elements are originally calculated with respect to a spherical harmonic oscillator s.p. basis.
By using the decomposition of the deformed s.p. wave function in Eq.~(\ref{4}), 
the two-body deformed wave function 
can be represented as:
\begin{eqnarray}
|p \bar{n}\rangle&=& \sum_{\eta_p\eta_n J}
F^{JK}_{p\eta_p n\eta_n}|\eta_p \eta_n,J K\rangle,
\label{decomp}
\end{eqnarray}
where 
$|\eta_p \eta_n, J K \rangle=\sum_{m_{p} m_{n}}
C^{JK}_{j_p m_{p} j_n m_{n} }|\eta_p m_{p}\rangle|\eta_n m_{n}\rangle$,
and  
$F^{JK}_{p\eta_p n\eta_n}= B^p_{\eta_p}B^n_{\eta_n}(-1)^{j_n-\Omega_{n}}C^{JK}_{j_p\Omega_{p} j_n-\Omega_{n}}$ 
is defined for the sake of simplicity
($(-1)^{j_n-\Omega_{n}}$ is the phase arising from the time-reversed states $|\bar{n}\rangle$).
The particle-particle $V_{p \bar {n},~{p'}\bar {n'}}$ and particle-hole $V_{p n',~p'n}$ 
interaction matrix elements in the representation (\ref{6}) 
for the QRPA matrices ${\cal A,\ B}$ (\ref{5}) in the deformed Woods-Saxon 
single-particle basis can then be given in terms of the spherical $G$~matrix elements as follows:
\begin{eqnarray}
V_{p \bar {n},~{p'}\bar {n'}}&= -&
2\sum_{J}\sum_{{\eta}_{p}{\eta}_{n}} \sum_{{\eta}_{p'}{\eta}_{n'}}
F^{JK}_{p\eta_p n\eta_n}F^{JK}_{p'\eta_{p'} {n'}\eta_{n'}}
G(\eta_p\eta_n\eta_{p'}\eta_{n'},J),\\
V_{p n',~p'n}&=&
2\sum_{J}\sum_{{\eta}_{p}{\eta}_{n}} \sum_{{\eta}_{p'}{\eta}_{n'}}
F^{JK'_{pn'}}_{p\eta_p {\bar n}'\eta_{n'}}
F^{JK'_{pn'}}_{p'\eta_{p'} \bar{n}\eta_{n}} G(\eta_p\eta_{n'}\eta_{p'}\eta_{n},J),
\label{10}
\end{eqnarray}
where $K'_{pn'}=\Omega_p+\Omega_{n'}=\Omega_{p'}+\Omega_n$.

The structure of the intermediate $| 0^+\rangle$ and $| 1^+\rangle$ states is only needed within the QRPA to calculate $2\nu\beta\beta$-decay NMEs $M^{2\nu}$~\cite{Sal09}, whereas
all possible $| K^\pi\rangle$ states are needed to construct the NMEs $M^{0\nu}$.

The matrix element $M^{2\nu}_{GT}$ is given within the QRPA in the intrinsic system by the following expression:
\begin{equation}
M^{2\nu}_{GT}= \sum_{K=0,\pm 1} \sum_{{m_i m_f}}
\frac{\langle 0^+_f| \bar\beta^-_{K} | K^+,m_f\rangle\langle K^+,m_f|K^+,m_i\rangle
\langle K^+,m_i| \beta^-_K | 0^+_i\rangle}{\bar\omega_{K,m_im_f}}.
\label{2}
\end{equation}
Instead of the usual approximation of the energy denominator in Eq.~(\ref{2}) as ${\bar\omega}_{K,m_im_f}=(\omega_{K,m_f} + \omega_{K,m_i})/2$ (see, e.g., Refs.~\cite{Sim03,Sim04}), here another prescription is used in which the whole calculated QRPA energy spectrum is shifted in such a way as to have the first calculated $1^+$ state exactly at the corresponding experimental energy. In this case the energy denominator in Eq.~(\ref{2}) acquires the form ${\bar\omega_{K,m_im_f}}=(\omega_{K,m_f} - \omega_{K,1_f} + \omega_{K,m_i}  - \omega_{K,1_i})/2+\bar\omega_{1^+_1}$, with $\bar\omega_{1^+_1}$ being the experimental excitation energy of the first $1^+$ state measured from the mean g.s. energy $(E_{0_i}+E_{0_f})/2$. 

The two sets of intermediate nuclear states generated from the
initial and final g.s. do not come out identical within the
QRPA. Therefore, the overlap factor of these states is introduced in Eq.~(\ref{2})~\cite{Sim03} as follows:
\begin{equation}
\langle K^+,m_f|K^+,m_i\rangle =
\sum_{l_i l_f}
~[X^{m_f}_{l_f K}X^{m_i}_{l_i K}-Y^{m_f}_{l_f K}Y^{m_i}_{l_i K}] 
\, {\cal R}_{l_f l_i}
\, \langle BCS_f|BCS_i\rangle.
\label{8}
\end{equation}
The factor ${\cal R}_{l_f l_i}$, which includes the overlaps of single particle 
wave functions of the initial and final nuclei is given  by:
\begin{eqnarray}
{\cal R}_{ll'}&=& \langle p \rho_p |p' {\rho}_{p'} \rangle(u^{(i)}_p u^{(f)}_{p'}+v^{(i)}_p v^{(f)}_{p'}) 
 \langle n \rho_n |n' {\rho}_{n'} \rangle(u^{(i)}_n u^{(f)}_{n'}+v^{(i)}_n v^{(f)}_{n'}),
\label{9}
\end{eqnarray} 
and the last term $\langle BCS_f|BCS_i\rangle$ in Eq.~(\ref{8}) corresponds to the overlap factor of the initial and final BCS vacua in the form given in Ref.~\cite{Sim03}.

The matrix element $M^{0\nu}$ is given within the QRPA in the intrinsic system by a sum of the partial amplitudes of transitions via all the intermediate states $K^\pi$:
\begin{equation}
M^{0\nu}=\sum_{K^\pi} M^{0\nu}(K^\pi)\ ; \ M^{0\nu}(K^\pi) = 
\sum_{\alpha} s^{(def)}_\alpha O_\alpha(K^\pi). \label{M0n}
\end{equation}
Here, we use the notation of Appendix B in Ref.~\cite{anatomy}, $\alpha$ stands for the set of four single-particle indices $\{p,p',n,n'\}$, 
and $O_\alpha(K^\pi)$ is a two-nucleon transition amplitude via the $K^\pi$ states
in the intrinsic frame:
\begin{equation}
O_\alpha(K^\pi)=\sum_{m_i,m_f}
\langle 0_f^+|c_{p}^\dagger c_{n}|K^\pi m_f\rangle 
\langle K^\pi m_f|K^\pi m_i\rangle
\langle K^\pi m_i|c^\dagger_{p'} c_{n'}|0_i^+\rangle .
\label{O}
\end{equation}
The two sets of intermediate nuclear states generated from the
initial and final g.s. (labeled by $m_i$ and $m_f$, respectively) 
do not come out identical within the
QRPA. A standard way to tackle this problem is to introduce in Eq.~(\ref{O}) the overlap factor of these states $\langle K^\pi m_f|K^\pi m_i\rangle$, whose representation is given below, Eq.~(\ref{overlap}).
Two-body matrix elements $s^{(def)}_\alpha$ of the neutrino potential in 
Eq.~(\ref{M0n}) in a deformed Woods-Saxon single-particle basis are decomposed over the the spherical harmonic oscillator ones according to 
Eqs.~(\ref{decomp},\ref{10}):
\begin{equation}
s^{(def)}_{pp'nn'}=
\sum_{J}\sum_{
\begin{array}{c}
\eta_p \eta_{p'}\\[-1pt]  \eta_n \eta_{n'}\end{array}}
F^{JK}_{p\eta_p n\eta_n}F^{JK}_{p'\eta_{p'}n'\eta_{n'}}s^{(sph)}_{\eta_p\eta_{p'} \eta_n\eta_{n'}}(J),
\end{equation}
\begin{eqnarray}
s^{(sph)}_{pp'nn'}(J)&=&\displaystyle \sum_{\mathcal J}
(-1)^{j_n + j_{p'} + J + {\mathcal J}} \hat{\mathcal J}
\left\{
\begin{array}{c c c}
j_p & j_n & J \\ j_{n'} & j_{p'} & {\mathcal J}
\end{array}
\right\} 
\langle p(1), p'(2); {\mathcal J} \| {\mathcal O_\ell}(1,2) \| n(1), n'(2); {\mathcal J} \rangle\,,
\end{eqnarray}
where $\hat{\mathcal{J}} \equiv \sqrt{2\mathcal{J}+1}$, and ${\mathcal O_\ell}(1,2)$ is the neutrino potential as a function of coordinates of two particles, with ${\ell}$ labeling its Fermi (F), Gamow-Teller (GT) and Tensor (T) parts. 

The particle-hole transition amplitudes in Eq.~(\ref{O}) can be represented in terms 
of the QRPA forward $X^m_{i K}$ and backward $Y^m_{i K}$  amplitudes along with 
the coefficients of the Bogoliubov transformation $u_\tau$ and $v_\tau$~\cite{Sal09}:
\begin{eqnarray}
\langle 0_f^+|c_{p}^\dagger c_{n}|K^\pi m_f\rangle&=&v_{p}u_{n}X^{m_f}_{pn,K^\pi}+u_{p}v_{n}Y^{m_f}_{pn,K^\pi},\nonumber\\
\langle K^\pi m_i|c^\dagger_p c_{n}|0_i^+\rangle&=&u_{p}v_{n}X^{m_i}_{pn,K^\pi}+v_{p}u_{n}Y^{m_i}_{pn,K^\pi}.\nonumber
\end{eqnarray}
The overlap factor in Eq.~(\ref{O}) can be written as:
\begin{eqnarray}
\langle K^\pi m_f|K^\pi m_i\rangle&=&\sum_{l_il_f}[X^{m_f}_{l_fK^\pi}X^{m_i}_{l_iK^\pi}-Y^{m_f}_{l_fK^\pi}Y^{m_i}_{l_iK^\pi}]
\mathcal{R}_{l_fl_i}\langle BCS_f|BCS_i\rangle
\label{overlap}
\end{eqnarray}
Representations for ${\cal R}_{l_fl_i}$  and the overlap factor $\langle BCS_f|BCS_i\rangle$ between the initial and final BCS vacua  are given in Ref.~\cite{Sim03}.

\subsection{Calculation results}

The NMEs $M^{2\nu}$ and $M^{0\nu}$ were calculated according to the above formalism in Refs.~\cite{Sim03,Sim04,Sal09,Fang10,Fang11}. These articles contain detailed description of the choice of the model parameters and comparison between different approximations.
Here we only briefly repeat the key points of the calculations.

Only quadrupole deformation is taken into account in the calculations~\cite{Sim03,Sim04,Sal09,Fang10,Fang11}.
The single-particle Schr\"odinger equation with the Hamiltonian of a deformed Woods-Saxon mean field is solved on the basis of an axially-deformed harmonic oscillator. 
Decomposition of the obtained deformed single-particle wave functions is performed over the spherical harmonic oscillator states within the seven lowest major shells.
The geometrical quadrupole deformation parameter $\beta_2$ of the deformed Woods-Saxon mean field is obtained by fitting the experimental deformation parameter $\beta= \sqrt{\frac{\pi}{5}}\frac{Q_p}{Z r^{2}_{c}}$, where $r_c $ is the charge rms radius and  $Q_p$ is the empirical intrinsic quadrupole moment. The experimental values of $\beta$
can be derived from the laboratory quadrupole moments measured by the Coulomb excitation reorientation technique, or from the corresponding $B(E2)$ values~\cite{ragha}. 
Experimental values extracted from the $B(E2)$ have smaller experimental eroors. But deformations extracted from the reorientation effect are in principle the better values, but have large errors. 
The fitted values of the parameter $\beta_2$ of the deformed Woods-Saxon mean 
field, which allow us to reproduce the experimental $\beta$, are listed in Table 1
of Ref.~\cite{Fang11}.
The spherical limit, i.e. $\beta_2=0$, is considered as well, 
to compare with the earlier results of Ref.~\cite{Rod05}. 

The nuclear Brueckner $G$~matrix, obtained by a solution of the Bethe-Goldstone equation with the Bonn-CD one boson exchange nucleon-nucleon potential, is used as a residual two-body interaction in Refs.~\cite{Sal09,Fang10,Fang11}. 
Then the BCS equations are solved to obtain the Bogoliubov coefficients, gap parameter and chemical potentials. 
To solve the QRPA equations, one has to fix the particle-hole $g_{ph}$ and particle-particle $g_{pp}$ renormalization factors of the residual interaction, Eqs.~(\ref{6}). 
A value of $g_{ph}=0.90$ was determined by fitting the experimental  position of the Gamow-Teller giant resonance (GTR) for $^{76}$Ge. The same value of $g_{ph}$ was then used for all nuclei in question, and led to a good fit to the experimental GTR energy for $^{150}$Nd, measured very recently~\cite{Guess11}.

The parameter $g_{pp}$ can be determined by fitting the experimental value of the $2\nu\beta\beta$-decay NMEs $M^{2\nu}_{GT}$ MeV$^{-1}$~\cite{Bar10} for each nucleus in question.
To account for the quenching of the axial-vector coupling constant $g_A$,  the quenched value $g_A^{eff}=0.75\cdot g_A$ was used in the calculation along with the bare value $g_A=1.25$. The quenching factor of 0.75 comes from a recent experimental measurement of GT strength distribution in $^{150}$Nd~\cite{Guess11}.
The two sets of the fitted values of $g_{pp}$ corresponding to the cases without or with quenching of $g_A$ are listed in Table 1 of Ref.~\cite{Fang11}.
Note, that this fitting procedure leads to realistic values $g_{pp}\simeq 1$.

\begin{table}[!t]
  \begin{center}
\caption{
Calculated NMEs $M'^{0\nu}$ and corresponding decay half-lives (assuming $m_{\beta\beta}$=50 meV) 
for $0\nu\beta\beta$-decays of $^{76}$Ge, $^{150}$Nd, $^{160}$Gd~\cite{Fang11},
and for $0\nu$ECEC of ${^{152}}$Gd, ${^{164}}$Gd, ${^{180}}$W \cite{ececfang}  within the deformed QRPA.
The results obtained for realistic deformations of the nuclei are labeled as ``def", whereas those 
obtained in the spherical limit, i.e. $\beta_2=0$, are labeled as ``sph". 
}
\begin{tabular}{lccccccc}\hline\hline
 Nuclear transition & $g_A^{eff}$ & & \multicolumn{2}{c}{sph ($\beta_2$=0)} & & \multicolumn{2}{c}{def}\\ \cline{4-5} \cline{7-8}
         &            & & \hspace{0.2cm} ${M'}^{0\nu}$ \hspace{0.2cm}  & \hspace{0.2cm} $T^{0\nu}_{1/2}$ [yr] \hspace{0.2cm}  
                      & & \hspace{0.2cm} ${M'}^{0\nu}$ \hspace{0.2cm}  & \hspace{0.2cm} $T^{0\nu}_{1/2}$ [yr] \hspace{0.2cm} \\ \hline
\multicolumn{8}{c}{$0\nu\beta\beta$-decay}\\   
${^{76}}$Ge$\rightarrow$${^{76}}$Se
 & 0.94 & &  4.10 & $9.4\times 10^{26}$ & &  4.00 & $9.8\times 10^{26}$ \\ 
            & 1.25 & &  5.30 & $5.6\times 10^{26}$ & &  4.69 & $7.2\times 10^{26}$ \\ 
${^{150}}$Nd$\rightarrow$${^{150}}$Sm 
 & 0.94 & &  4.52 & $2.3\times 10^{25}$ & &   2.55 & $7.1\times 10^{25}$ \\ 
 & 1.25 & &  6.12 & $1.2\times 10^{25}$ & &   3.34 & $4.1\times 10^{25}$ \\
${^{160}}$Gd$\rightarrow$${^{160}}$Dy
 & 0.94 & &       &                     & &    3.76 & $2.3\times 10^{26}$ \\
\multicolumn{8}{c}{$0\nu$ECEC}\\     
${^{152}}$Gd$\rightarrow$${^{152}}$Se (KL$_1$)
 & 1.269 & &  7.50 & ($8.7\times 10^{27}$, $8.9\times 10^{29}$) & &  3.23 & ($4.7\times 10^{28}$, $4.8\times 10^{29}$) \\
${^{164}}$Er$\rightarrow$${^{164}}$Dy (L$_1$L$_1$)
 & 1.269 & &  7.20 & ($1.0\times 10^{32}$, $1.1\times 10^{32}$) & &  2.64 & ($7.5\times 10^{32}$, $8.4\times 10^{32}$) \\
${^{180}}$W$\rightarrow$${^{180}}$Hf (KK)
 & 1.269 & &  6.22 & ($1.4\times 10^{30}$, $2.0\times 10^{30}$) & &  2.05 & ($1.3\times 10^{31}$, $1.8\times 10^{31}$) \\
\hline\hline
\end{tabular}
\label{MEres}
\end{center}
\end{table}

Having solved the QRPA equations, the two-nucleon transition amplitudes (\ref{O}) are calculated and, by combining them with the two-body matrix elements of the neutrino potential, the total $0\nu\beta\beta$ NMEs $M^{0\nu}$ (\ref{M0n}) is formed. 
Such a computation is rather time consuming since numerous programming loops are needed to calculate the decompositions of the two-body matrix elements in the deformed basis over the spherical ones. Therefore, to speed up the calculations the mean energy of 7 MeV of the intermediate nuclear excitation energies  is used in the neutrino propagator.
The effects of the finite nucleon size and higher-order weak currents were taken into account. The two-nucleon short-range correlations (SRC) were treated in an extended Brueckner theory (CCM)  in a modern self-consistent way, see~\cite{src09} and Sect.II, that leads to a change in the NMEs $M^{0\nu}$  only by a few percents, much less than the traditional Jastrow-type representation of the SRC does. 

An important cross-check of the calculations is provided by a comparison of the present results in the spherical limit 
with the previous ones of Refs.~\cite{Rod05,anatomy,src09}. 
Though formally the adiabatic Bohr-Mottelson approximation is not applicable in the limit of vanishing deformation, it is easy to see that the basic equations~(\ref{M0n})--(\ref{overlap}) do have the correct spherical limit. Details of such a comparison can be found in Ref.~\cite{Fang11}, and an excellent agreement between the NMEs calculated by the genuine spherical code and the deformed code in the spherical limit was found.
Also, the partial contributions $M^{0\nu}(K^\pi)$ of different intermediate $K^\pi$ states to $M^{0\nu}$ for the decay $^{150}$Nd$\rightarrow ^{150}$Sm were analyzed in 
Ref.~\cite{Fang11}.

The final results for the NMEs, corresponding to the modern self-consistent treatment of the SRC~\cite{src09}), for $0\nu\beta\beta$ decays $^{76}$Ge$\rightarrow ^{76}$Se, $^{150}$Nd$\rightarrow ^{150}$Sm, $^{160}$Gd$\rightarrow ^{160}$Dy
are listed in Table~\ref{MEres}. 
As explained in Ref.~\cite{Fang11}, the difference between the spherical and deformed results mainly come from the BSC overlap between the ground states of the initial and final nuclei. The strongest effect of deformation on $M^{0\nu}$ (the suppression by about 40\% as compared to our previous QRPA result obtained with neglect of deformation) is found  in the case of $^{150}$Nd. This suppression can be traced back to a rather large difference in deformations of the ground states of $^{150}$Nd and $^{150}$Sm. As for the $g_{pp}$ dependence of the $0\nu\beta\beta$-decay NMEs, 
it is much less pronounced than the dependence of the amplitude of $2\nu\beta\beta$ decay.
A marked reduction of the total $M'^{0\nu}$ for the quenched value of $g_A$
can be traced back to a smaller prefactor $(g_A/1.25)^2$ in the definition of $M'^{0\nu}$~(\ref{nmep}).

In Table~\ref{tab:3} the NMEs $M^{0\nu}$ for $^{150}$Nd calculated by other approaches are compared. 
The NMEs $M^{0\nu}$ for $^{150}$Nd, obtained within the state-of-the-art QRPA approach that accounts for nuclear deformation~\cite{Fang11}, 
compares well with the results of the IBM~\cite{Bar09} and PHFB~\cite{Hir08}.
The calculated $0\nu\beta\beta$-decay half-life $T^{0\nu}_{1/2}$ corresponding to the Majorana neutrino mass $\langle m_{\beta\beta} \rangle$ = 50 meV seems to be short enough
to hope that the SNO+ experiment will be able to approach the inverse hierarchy of the neutrino mass spectrum.

\begin{table}[!t]
\caption{
The matrix elements $M^{0\nu}$ for the $0\nu\beta\beta$ decay $^{150}$Nd$\rightarrow ^{150}$Sm calculated in different models. 
The corresponding half-lives $T^{0\nu}_{1/2}$ (in years) for an assumed effective Majorana neutrino mass $\langle m_{\beta\beta} \rangle$ = 50 meV are also shown.
}
\begin{tabular}{cccccc} 
\hline
Method
&  {def. QRPA~\cite{Fang11}}
&  {pseudo-SU(3)~\cite{Hir95}}
&  {PHFB~\cite{Hir08}}
&  {IBM~\cite{Bar09}}
&  {EDF~\cite{Rodri10}}
\\
\hline  
$M^{0\nu}$   
& $2.95\pm 0.4$ & 1.57 & $3.24\pm 0.44$ & 2.32 & 1.71\\
$T^{0\nu}_{1/2}$ ($10^{25}$ y)
& $5.6\pm 1.5$ & 18.7 & $4.6\pm 1.2$ & 8.54 & 16.5\\
\hline
\end{tabular} 
\label{tab:3}
\end{table}

\section{On the possibility to measure Fermi $0\nu\beta\beta$-decay nuclear matrix element}

Although there has been great progress in the calculations of the NME $M^{0\nu}$ over the last decade, but still there can be a substantial scatter in the calculated $M^{0\nu}$ by different groups. Even more striking, up to a factor of 5, can be the difference in the Fermi part $M^{0\nu}_F$ of the total $M^{0\nu}$.

Therefore, it would be very important to find  a possibility to determine $M^{0\nu}$ experimentally. Partial one-leg transition amplitudes to the intermediate $1^+$ states have been measured by charge-exchange reactions in many nuclei (see~\cite{frekers} and references therein), thereby providing important spectroscopic information. 
However, an attempt to reconstruct the nuclear amplitude $M^{2\nu}$ of two-neutrino $\beta\beta$ decay from the measured amplitudes suffers from large inherent uncertainties stemming from unknown relative phases of different intermediate-state contributions.  Thus, only if a transition via a single intermediate $1^+$ state dominates $M^{2\nu}$, $M^{2\nu}$ can consistently be determined.
Trying the same way to reconstruct $M^{0\nu}$ seems even more hopeless, since  
many intermediate states of different multipolarities (with a rather moderate contribution of the $1^+$ states)
are virtually populated in the $0\nu\beta\beta$ decay due to a large momentum of the exchanged virtual neutrino. In addition, the transition operators in a charge-exchange reaction and $0\nu\beta\beta$ decay become more and more different for higher spins of the intermediate states.

A proposal suggesting a way of a direct measurement of $M^{0\nu}_F$ was put forward in a recent work~\cite{Rod09}. It exploits
the similarity between the Fermi part of the neutrino potential in $0\nu\beta\beta$ decay and the radial dependence of the Coulomb interaction. The latter is well-known to be the leading source of the isospin breaking in nuclei~\cite{auer72,auer83}. As shown in Ref.~\cite{Rod09}, the Fermi matrix element $M_{F}^{0\nu}$ can be related to the Coulomb mixing matrix element between the ideal double isobaric analog state (DIAS) of the ground state (g.s.) of the initial nucleus~\footnote{This ideal DIAS would be an exact nuclear state if the isospin symmetry were exact.} and the g.s. of the final nucleus. As a result of the Coulomb mixing,
the single Fermi transition matrix element $\langle 0_f | \hat T^{-} | IAS \rangle$ between the isobaric analog state (IAS) of the g.s. of the initial nucleus and the g.s. of the final nucleus becomes non-zero.
Thus, having measured this single Fermi transition matrix element $\langle 0_f | \hat T^{-} | IAS \rangle$ by charge-exchange reactions, the $0\nu\beta\beta$-decay matrix element $M_{F}^{0\nu}$ can be reconstructed. 

Of course, by measuring only $M_{F}^{0\nu}$ one would not get the total matrix element $M^{0\nu}$ but rather its sub-dominant part contributing about 20--30\% to 
$M^{0\nu}$. However, knowledge of $M_{F}^{0\nu}$ itself brings a very important piece of information. For instance, 
it can help to discriminate between  different nuclear structure models in which calculated $M_{F}^{0\nu}$ may differ by as much as a factor of 5.
More importantly, the ratio $M_{F}^{0\nu}/M_{GT}^{0\nu}$ may be more 
reliably calculable in different models than $M_{F}^{0\nu}$ and $M_{GT}^{0\nu}$ separately. 
Simple arguments put forward in Ref.~\cite{Rod09} showed that  
an estimate $M_{GT}^{0\nu}/M_{F}^{0\nu}\approx -2.5$ should hold in a realistic calculation (QRPA results~\cite{Rod05,anatomy,src09} do agree with this simple estimate). 

The master relation, derived in Ref.~\cite{Rod09} in the closure approximation~\footnote{Using closure of the states of the intermediate nucleus $_{Z+1}^{\phantom{+2} A} {\mathrm{El}}_{N-1}$ which are virtually excited in $\beta\beta$-decay would be an exact procedure if there were no energy dependence in the $0\nu\beta\beta$ transition operator. A weak energy dependence of the operator leads in reality to a ``beyond-closure'' correction to the total $M^{0\nu}$ which does not exceed 10\%.} by making use of the isospin symmetry of strong interaction $\hat H_{str}$, represents 
the matrix element $M_{F}^{0\nu}$ in the form of an energy-weighted double Fermi transition matrix element:
\be
M^{0\nu}_F = - \frac{2}{e^2} 
\sum_s \bar\omega_s \langle 0_f | \hat T^{-} |0^+_s \rangle  \langle 0^+_s | \hat T^{-} |0_i\rangle.
\label{MFtot}
\ee
Here, $\hat T^{-}=\sum_{a}\tau_a^{-}$ is the isospin lowering operator,
the sum runs over all $0^+$ states of the intermediate nucleus $_{Z+1}^{\phantom{+2} A} {\mathrm{El}}_{N-1}$, 
$\bar\omega_s=E_s-(E_{0_i}+E_{0_f})/2$
is the excitation energy of the intermediate state $s$ relative to the mean energy of g.s. of the initial and final nucleus.  

To account for the isospin-breaking part of $\hat H_{str}$, 
the r.h.s. of Eq.~(\ref{MFtot}) should be slightly modified.
It is well known that the isospin-breaking terms in $\hat H_{str}$ are in fact fairly small, at the level of 2\%--3\%~\cite{auer72,auer83}, and we  
arrived in Ref.~\cite{Rod09} at the conclusion that the contribution of this source of the isospin non-conservation to Eq.~(\ref{MFtot}) is about 20--30 \% of that caused by the Coulomb interaction.

As argued in Ref.~\cite{Rod09}, the expression (\ref{MFtot}) in the leading order of the Coulomb mixing must be dominated by the amplitude of the double Fermi transition from the initial g.s. via its IAS into the final g.s.:
\begin{equation}
M^{0\nu}_F \approx - \frac{2}{e^2}\,\bar\omega_{IAS} 
\langle 0_f | \hat T^{-} |IAS \rangle  \langle IAS | \hat T^{-} |0_i\rangle ,
\label{MFappr}
\end{equation}
Here, the second Fermi transition amplitude is non-vanishing due to an admixture of the ideal double IAS (DIAS) wave function
$|DIAS\rangle=\frac{ (\hat T^{-})^2}{\sqrt{4T_0(2T_0-1)}} |0_i^+\rangle$
in the g.s. of the final nucleus: 
$\langle 0_f |\hat T^{-}| IAS\rangle = \langle 0_f | DIAS\rangle \langle DIAS |\hat T^{-}| IAS\rangle$, and $T_0=(N-Z)/2$ is the isospin of the g.s. of the initial nucleus. 

In Eq.~(\ref{MFappr}), the first-leg matrix element $\langle IAS | \hat T^{-} | 0_i \rangle \approx \sqrt{2T_0}=\sqrt{N-Z}$ and the IAS energy $\omega_{IAS}$ are very accurately known.
Thus, the total $M^{0\nu}_F$ can be reconstructed according to Eq.~(\ref{MFappr}), 
if one is able to measure the $\Delta T=2$ isospin-forbidden matrix element 
$\langle IAS | \hat T^{+} | 0_f \rangle$, for instance in charge-exchange reactions of the $(n,p)$-type. 

From the value of $M^{0\nu}_F$ calculated in a model, the magnitude of the matrix element $\langle IAS | \hat T^{+} | 0_f \rangle$ can be estimated by using a transformed version of Eq.~(\ref{MFappr}):
\be 
\langle IAS | \hat T^{+} | 0_f \rangle=  - \frac{e^2 M^{0\nu}_F}{2 \bar\omega_{IAS} \sqrt{N-Z} }.
\label{M}
\ee
Using recent QRPA calculation results for $M^{0\nu}_F$~\cite{Rod05}, this matrix element can roughly be estimated as $\langle IAS | \hat T^{+} | 0_f \rangle \sim 0.005$,  
i.e. about thousand times smaller than the first-leg matrix element $\langle IAS | \hat T^{-} | 0_i \rangle$. 
This strong suppression of $\langle IAS | \hat T^{+} | 0_f \rangle$ reflects the smallness of the isospin-breaking effects in nuclei. 

The IAS has been observed as a pronounced and extremely narrow resonance, and its various features have well been studied 
by means of $(p,n)$, ($^3$He,$t$) and other charge-exchange reactions on the g.s. of a mother nucleus. In this case the reaction cross-section at the zero scattering angle can be shown to be proportional to a large Fermi matrix element $\langle IAS | \hat T^{-} | 0_i \rangle \approx \sqrt{N-Z}$~\cite{tad87}.
Extraction of a strongly suppressed matrix element $\langle IAS | \hat T^{+} | 0_f \rangle$ from a tiny cross-section of the $(n,p)$-type reactions on the final nucleus might only be possible if there exists a similar proportionality in the $(n,p)$ channel.

Therefore, a detailed realistic analysis of the corresponding reaction mechanism is needed to assess the possibility of extraction of the matrix element $\langle IAS | \hat T^{+} | 0_f \rangle$ from the corresponding reaction cross-sections.

A first preliminary assessment of the $(n,p)$ reaction at the zero scattering angle was done in Ref.~\cite{Rod10}. The IAS was treated as a single, well-isolated, state as it appears in rather light nuclei. 
In such a case the Coulomb mixing could be treated perturbatively that significantly simplified the consideration.  
In heavier nuclei the spread of the IAS becomes rather significant and should be taken into consideration. 

As argued in Ref.~\cite{Rod10}, the isospin of the projectile should not be larger than $T=1/2$. Indeed, the main components of the wave functions $| 0_f \rangle$ and $| IAS \rangle$ have the total isospin different by two units. Therefore, already for a projectile with isospin $T=1$ a common entrance and exit isospin channel exists arising from a recoupling of the isospin $T=1$ of the projectile with the main components of the wave functions of the target and daughter nuclei. In such a case extraction of the information about small isospin impurities from the corresponding reaction cross-section seems barely possible.

Thus, the only probes which seem feasible are of the isospin $T=1/2$ ($(n,p)$, ($t,^3$He),
 \dots reactions). However, it is still not guaranteed that the reaction cross-section $\sigma(0_f^+\to IAS)$ for these probes is proportional to a strongly suppressed matrix element $\langle IAS | \hat T^{+} | 0_f \rangle$, since the other isospin impurities in the wave functions $| 0_f \rangle$ and $| IAS \rangle$ may have a larger effect on the reaction cross-section.

A preliminary assessment of the $(n,p)$ reaction at the zero scattering angle in the aforementioned perturbative mixing approximation was performed in Ref.~\cite{Rod10} for an intermediate-mass $\beta\beta$-decaying nucleus $^{82}$Se. In fact, it was shown that the tiny cross-section $\sigma_{np}(0_f^+\to IAS)$ is indeed dominated by the admixture of the DIAS in the g.s. of the final nucleus. However, the spread of the IAS of $^{82}$Se is rather significant that may question the perturbative treatment of the isospin mixing.

A $\beta\beta$-decaying nucleus in which such a perturbative treatment may be justified in reality is $^{48}$Ca. This case was considered in detail in Ref.~\cite{Rod11}.

The IAS of $^{48}$Ca is a state with $J^\pi=0^+, T=4, T_z=3$ at the excitation energy of $E_x=$6.678 MeV ($\bar\omega_{IAS}\approx $8.5 MeV) in $^{48}$Sc. It lies under the threshold of particle emission and with almost 100\% probability decays to $1^+$ state at $E_x=$2.517 MeV by the emission of a $\gamma$-quantum with $E_\gamma$=4.160 MeV~\cite{Folk75}. This $\gamma$-decay energy is much higher than the $\gamma$-decay energies from $0^+$ states with normal isospin $T=T_z=3$ surrounding the IAS (which decay by a cascade), and could be used as a unique experimental tag telling that the IAS indeed was excited in a reaction.

There are strong arguments that the IAS of $^{48}$Ca must be a single state without fragmentation. 
The state-of-the-art measurement of $^{48}$Ca($^{3}$He,$t$)$^{48}$Sc(IAS) reaction~\cite{Gr07} does in fact contribute to clarification of this issue as discussed below.

Fragmentation of the IAS may occur only if there are several $0^+$ states with the normal isospin around the IAS to which the IAS may strongly couple. 
In other words, the total number of the $0^+$ states within the IAS spreading width 
$\Gamma^\downarrow_A$ must be greater than one (for nuclei around $A=50$ \ \ $\Gamma^\downarrow_A$  is typically about few keV).
Based on the back-shifted Fermi-gas model~\cite{Dilg73},
the mean level spacing  between the $0^+$ states of the normal isospin in the vicinity of the IAS in $^{48}$Sc was estimated in Ref.~\cite{Rod11} to be about 50--70 keV.
Then, if the IAS were essentially spread over those $0^+$ states, the experiment~\cite{Gr07} would have been able to resolve components of the IAS fine structure. The fact that no fine structure was observed can easily be understood from a comparison of a typical $\Gamma^\downarrow_A$ of the order of few keV with a much larger mean level spacing.

The cross-section for the reaction $^{48}$Ti(n,p)$^{48}$Sc(IAS) was estimated in Ref.~\cite{Rod11} to be $\frac{d^2\sigma_{np}}{d\Omega dE}\approx 20$ nb/(sr MeV) for the energy of the incident neutrons around 100 MeV.
Note, that by choosing a smaller neutron incident energy this estimate can further be improved by a factor of 2--3, due to the increasing Fermi unit cross-section~\cite{tad87}.
The Coulomb mixing of the IAS with the isovector monopole resonance was estimated to  modify the above value of $\frac{d^2\sigma_{np}}{d\Omega dE}$ by few percents, and therefore can be neglected.


\section{The resonant neutrinoless double electron capture}

The resonant $0\nu$ECEC (neutrinoless double electron capture) was considered as process, which might prove 
the Majorana nature of neutrinos and the violation of the total lepton number,
by Winter \cite{WINTER} already in 1955. The possibility of a resonant enhancement 
of the $0\nu$ECEC in case of a mass degeneracy between the initial and final atoms 
was pointed out  by Bernab\'eu, De Rujula, and Jarlskog as well as by Vergados 
about 30 years ago \cite{Ver83,BeRuJar83}. They estimated the half-life of the process 
by introducing different simplifications: i) non-relativistic atomic wave functions 
at nuclear origin; ii) qualitative evaluation of NME of the process;
iii) the degeneracy parameter $\Delta = M_{A,Z} - M_{A,Z - 2}^{**}$ was assumed
to be within the range (0,10) keV representing the accuracy of atomic mass 
measurement at that time. $M_{A,Z}$ and $M_{A,Z - 2}^{**}$ are masses of the initial 
and final excited atoms, respectively. A list of promising isotopes based 
on the degeneracy requirement associated with arbitrary nuclear excitation 
and on the natural abundance of daughter atom was presented. 

In 2004 Sujkowski and Wycech \cite{sujwy} and Lukaszuk et al. \cite{lukas} analyzed 
the $0\nu$ECEC process for  nuclear $0^+ \to 0^+$ transitions accompanied by a photon 
emission in the resonance and non-resonance modes. By assuming $|m_{\beta\beta}|=1$ eV 
and 1 $\sigma$ error in the atomic mass determination the resonant $0\nu$ECEC rates 
of six selected isotopes were calculated by considering the perturbation theory approach. 

In 2009 a new theoretical approach to the $0\nu$ECEC, a unified description of 
the oscillations of stable and quasistationary atoms, was developed by \v Simkovic 
and Krivoruchenko \cite{mik11,SIM08}. A comprehensive theoretical study of 
this process for the light Majorana neutrino mass mechanism was performed \cite{mik11}. 
It was shown that effects associated with the relativistic structure of the electron 
shells reduce the $0\nu$ECEC half-lives by almost one order of magnitude and that 
the capture of electrons from the $n p_{1/2}$ states is only moderately suppressed 
in comparison with the capture from the $n s_{1/2}$ states unlike in the non-relativistic 
theory. Selection rules for associated nuclear transitions were presented saying 
that a change in  the nuclear spin $J \ge 2$ are strongly suppressed. New transitions 
due to the violation of parity in the $0\nu$ECEC process were proposed, e.g., nuclear 
transitions $0^+ \rightarrow   0^\pm, 1^\pm$ are compatible with a mixed capture of s- and 
p-wave electrons. Based on the most recent data and realistic evaluation of 
the decay half-lives,  a complete list of the most perspective isotopes for which
the $0\nu$ECEC capture may have the resonance enhancement was provided.
for further experimental study. It inludes $^{96}Ru$, $^{106}Cd$, $^{124}Xe$,
$^{136}Ce$, $^{152}Gd$, $^{156}Dy$, $^{164}Er$, $^{168}Yb$,
$^{180}W$, $^{184}Os$ and $^{190}Pt$ \cite{mik11}. 
By assuming $|m_{\beta\beta}|=50$ meV  and an appropriate 
value of NME, half-lives of some of the isotopes were found to be as low as $10^{25}$ 
years in the unitary limit. It is about one order of magnitude shorter than the 
$0\nu\beta\beta$ half-life of $^{76}$Ge for the same value of the 
effective mass of Majorana neutrinos.  

The inverse value of the half-life of resonant neutrinoless double electron capture
\begin{equation} 
\frac{\ln{2}}{T^{0\nu\mathrm{ECEC}}_{1/2}(J^\pi)} = \frac{\left| V_{a b}(J^\pi) \right|^2} 
{\Delta^2 + \frac{1}{4} \Gamma^2_{ab}} \Gamma_{ab}
\label{e:5} 
\end{equation} 
where $J^\pi$ denotes angular momentum and parity of final nucleus. 
The degeneracy parameter can be expressed as
\begin{equation}
\Delta = M_{A,Z} - M_{A,Z - 2}^{**} = Q - B_{ab} -E_\gamma,
\end{equation}
where  $Q$ stands for a difference between the initial and final atomic masses 
in ground states and $E_\gamma$ is an excitation energy of the daughter nucleus. 
$B_{ a b} = E_{a} + E_{b} + E_C $ is the energy of two electron holes, 
whose quantum numbers $(n, j, l)$ are denoted by indices $a$ and $b$
and $E_C$ is the interaction energy of the two holes.
The binding energies of single electron holes $ E_{a}$ are 
known  with accuracy with few eV \cite{LARKINS}. The 
width of the excited final atom with the electron holes is given by
\begin{equation}
\Gamma_{ab} = \Gamma_{a} + \Gamma_{b} + \Gamma^*.
\end{equation}
Here, $\Gamma_{a, b}$ is one-hole atomic width 
and $\Gamma^*$ is the de-excitation width of daughter nucleus, which can be neglected.
Numerical values of $\Gamma_{ab}$ are about up to few tens eV.
By factorizing the electron shell structure and nuclear matrix element 
for lepton number violating amplitude associated with nuclear transitions 
$0^+ \rightarrow J^\pi = 0^{\pm 1}, 1^{\pm 1}$  one gets
\begin{eqnarray}
V_{a b} (J^{\pi}) = 
\frac{1}{4 \pi}~ G^2_{\beta}m_e \eta_{\nu}  \frac{(g^{eff}_A)^2}{R} 
<F_{a b}> M^{0\nu ECEC}(J^\pi).
\label{potential}
\end{eqnarray}
Here, $<F_{a b}>$ is a combination of averaged upper and lower bispinor 
components of the atomic electron wave functions \cite{mik11}
and  $M^{0\nu ECEC}(J^\pi)$ is the nuclear matrix element. We note that 
by neglecting the lower bispinor components $M^{0\nu ECEC}(0^+)$ takes 
the form of the $0\nu\beta\beta$-decay NME for ground state to
ground state transition after replacing isospin operators $\tau^-$ 
by $\tau^+$. R is the nuclear radius and $g_A$ is the axial-vector 
coupling constant. 

The probability of the $0\nu$ECEC is increased by many orders of magnitude
provided the resonance condition  is satisfied within a few tens of 
electron-volts. For a long time  there was no way to  
identify promising isotopes for experimental search of $ 0\nu$ECEC, 
because of poor experimental accuracy of measurement of $Q$-values 
of the order of 1 - 10 keV for medium heavy nuclei. Progress in precision 
measurement of atomic masses with Penning traps \cite{penning1,penning2,penning3} 
has revived the interest in the old idea on the resonance $0\nu$ECEC.  
The accuracy of $Q$-values at around 100 eV was achieved. The estimates of 
the $0\nu$ECEC half-lives were recently improved by more accurate  
measurements of $Q$-values for ${^{74}}$Se \cite{kolhinen10,mount10},
${^{96}}$Ru \cite{elis2}, ${^{106}}$Cd \cite{smorra11,elis4},
${^{102}}$Pd \cite{elis4}, ${^{112}}$Sn \cite{RAKH09}, ${^{120}}$Te \cite{SCIE09},
${^{136}}$Ce \cite{kolhinen11}, ${^{144}}$Sm \cite{elis4}, ${^{152}}$Gd \cite{elis1}, 
${^{156}}$Dy \cite{elis3}, ${^{162}}$Er \cite{elis2}, ${^{164}}$Er \cite{elis5}, 
${^{168}}$Yb \cite{elis2} and ${^{180}}$W \cite{droese11}. It allowed
to exclude some of isotopes from the list of the most promising 
candidates (e.g., $^{112}$Sn, $^{164}$Er, ${^{180}W}$W) for searching the $0\nu$ECEC. 

Among the promising isotopes, $^{152}$Gd has likely resonance transitions to 
the $0^+$ ground states of the final nucleus as it follows from improved 
measurement of Q-value for this transition with accuracy of about 100 eV  \cite{elis1}.
A detailed calculation of the $0\nu$ECEC of $^{152}\mathrm{Gd}$ was performed in \cite{ececfang}
(see Table \ref{MEres}).
The atomic electron wave functions were treated in the relativistic 
Dirac-Hartree-Fock approximation \cite{MAWA1973}.  The NME for ground state
to ground state transition ${^{152}\mathrm{Gd}}\rightarrow {^{152}\mathrm{Sm}}$ was 
calculated within the proton-neutron deformed QRPA with a realistic residual interaction 
\cite{elis1}. For the favored capture of electrons from K and L shells in the
case of ${^{152}}$Gd the $0\nu$ECEC half-life is in the range
$4.7 \times 10^{28}$ - $4.8\times 10^{29}$ years. This transition is still rather
far from the resonant level. Currently, the  $0\nu$ECEC half-life of $^{152}$Gd 
is 2-3 orders of magnitude longer than the half-life of $0\nu\beta\beta$ decay 
of $^{76}$Ge corresponding to the same value of $|m_{\beta\beta}|$ and is 
the smallest known half-life among known $0\nu$ECEC.

The resonant $0\nu$ECEC  has some  important advantages with respect to experimental 
signatures and background conditions.  The ground state to ground state resonant 
$0\nu$ECEC transitions can be detected by monitoring the X rays or Auger electrons 
emitted from excited electron shell of the atom. This can be achieved, e.g.,  
by calorimetric measurements. The de-excitation of the final excited nucleus proceeds 
in most cases through a cascade of easy to detect  rays. A coincidence setup can cut 
down any background rate right from the beginning, thereby requiring significantly less 
active or passive shielding. A clear detection of these $\gamma$ rays 
would already signal the resonant $0\nu$ECEC without any doubt, as there are no 
background processes feeding those particular nuclear levels. We note that 
standard model allowed double electron capture with emission of two neutrinos, 
\begin{equation}
e^-_b + e^-_b + (A,Z) \rightarrow (A,Z-2)^{**} + \nu_e + \nu_e,
\end{equation}
is strongly suppressed due to almost vanishing phase space \cite{BeRuJar83}.

Till now, the most stringent limit on the resonant $0\nu$ECEC were established 
for $^{74}$Se \cite{barab1,FPS11}, $^{106}$Cd \cite{rukh11} and $^{112}$Sn \cite{barab2}.
However, following recent theoretical analysis \cite{mik11} none of these resonant 
$0\nu$ECEC transitions is favored in the case of light neutrino mass mechanism. 
The ground state of $^{74}$Se is almost degenerate to the second excited
state at 1204 keV in the daughter nucleus $^{74}$Ge,
which is a $2^+$ state \cite{frekdeg} and is disfavored by the selection rule \cite{mik11}. 
The TGV experiment situated in Modane established limit on the $0\nu$ECEC half-life of $1.1\times 10^{20}$ years 
\cite{rukh11}. Subject of interest was the $0\nu$ECEC resonant decay mode of $^{106}$Cd (KL-capture) 
to the excited 2741 keV state of $^{106}$Pd. For a long time the spin value of this final state 
was unknown  and it was assumed to be $J^\pi =(1, 2)^+$. However, a new value for the spin of 
the 2741 keV level in $^{106}$Pd is $J = 4^+$ and this transition is disfavored again 
due to selection rule. A search for the resonant $0\nu$ECEC in ${^{106}}$Cd was carried out 
also at the Gran Sasso National Laboratories with the help of a ${^{106}}CdWO_4$ 
crystal scintillator (215 g) enriched in ${^{106}}$Cd up to 66\% \cite{belli12}. It was found that 
the resonant $0\nu$ECEC to the 2718 keV ($J^\pi$ is unknown), 2741 keV ($J^\pi = 4^+$) 
and 2748 ($J^\pi = (2,3)^-$)  keV excited states of 
${^{106}}$Pd are restricted to $T^{0\nu ECEC}_{1/2}\ge 4.3\times 10^{20}$ yr (KK-capture), 
$T^{0\nu ECEC}_{1/2}\ge 9.5\times 10^{20}$ yr (KL$_1$-capture) and
$T^{0\nu ECEC}_{1/2}\ge 4.3\times 10^{20}$ yr (KL$_3$-capture), respectively. 
We note that the 2718 excited state $\gamma$ decays by 100\% into the 3+ state 
at 1557.68 keV state, which again excludes  a possibility of J = 0, 1 for this state.
Further, we already mentioned above that
a new mass measurement \cite{RAKH09} has excluded a complete mass degeneracy for a $^{112}$Sn 
decay and has therefore disfavored significant resonant enhancement of the $0\nu$ECEC
mode for this transition. Recently, a first bound on the resonant $0\nu$ECEC half-life of $^{136}$Ce
of about $1.1\times 10^{15}$ years  was measured \cite{belli09}.

\section{Summary}

Many new projects for measurements of the $0\nu\beta\beta$-decay have been proposed, 
which hope to probe effective neutrino mass $m_{\beta\beta}$ down to 10-50 meV. 
An uncontroversial detection of the $0\nu\beta\beta$‐decay will prove 
the total lepton number to be broken in nature, and neutrinos to be 
Majorana particles. There is a general consensus that 
a measurement of the $0\nu\beta\beta$-decay in one isotope does not allow 
us to determine the underlying physics mechanism. 
It is very desired that experiments involving as 
many different targets as possible to be pursued. There is also a revived 
interest to theoretical and experimental study of the resonant 
$0\nu$ECEC (neutrinoless double electron capture), which can probe the Majorana nature of neutrinos and 
the neutrino mass scale as well. The $0\nu$ECEC half-lives might be 
comparable to the shortest half-lives of the $0\nu\beta\beta$ decays 
of nuclei provided the resonance condition is matched with an accuracy 
of tens of electron-volts. There is a lot of theoretical and experiment
effort to determine the best $0\nu$ECEC candidate. 

Nuclear matrix elements of these two lepton number violating processes 
need to be evaluated with uncertainty of less than 30\% 
to establish the neutrino mass spectrum and CP violating phases. 
Recently, there has been significant progress in understanding the source 
of the spread of calculated NMEs. Nevertheless, there is no consensus among 
nuclear theorists about their correct values, and corresponding uncertainty. 
The improvement of the calculation of the nuclear matrix elements is 
a very important and challenging problem. We presented improved 
calculation of the $0\nu\beta\beta$-decay and $0\nu$ECEC NMEs, which 
includes a consistent treatment of the two-nucleon short-range
correlations and deformation effects. In addition, a possibility to
measure the $0\nu\beta\beta$-decay NME was addressed.

\acknowledgments 
 
This work was supported in part by the Deutsche Forschungsgemeinschaft  
within the project "Nuclear matrix elements of Neutrino Physics and Cosmology" 
FA67/40-1, the VEGA Grant agency of the Slovak Republic under the contract No.~1/0639/09 
and by the grant of the Ministry of Education and Science of the Russian Federation 
(contract 12.741.12.0150).


\begin{thebibliography}{99}


\bibitem{AEE08} F.T. Avignone, S.R. Elliott and J. Engel, 
      Rev. Mod. Phys. {\bf 80}, 481 (2008).  

\bibitem{ves12} J.D. Vergados, H. Ejiri, and F. \v Simkovic, arXiv:1205.0649 [hep-ph].

\bibitem{mik11} M.I. Krivoruchenko, F. \v Simkovic, D. Frekers, A. Faessler, 
   Nucl. Phys. A {\bf 859}, 140 (2011). 

\bibitem{zdes} V.I. Tretyak and Yu.G. Zdesenko, 
      At. Dat. Nucl. Dat. Tabl. {\bf 80}, 83 (2002).

\bibitem{bau99} L. Baudis {\it et al.} (The Heidelberg-Moscow collaboration),
     Phys. Rev. Lett. {\bf 83}, {41} (1999).

\bibitem{tre11} V.I. Tretyak (The NEMOIII collaboration), 
     AIP Conf. Proc. {\bf 1417}, 125 (2011).

\bibitem{te130} C. Arnaboldi et al. (The CUORE collaboration),
     Phys. Lett. B {\bf 584}, 260 (2004).

\bibitem{kamlandzen} A. Gando {\it et al.} (The KamLAND-Zen collaboration), 
     arXiv:1201.4664 [hep-ex].

\bibitem{evidence2} H.V. Klapdor-Kleingrothaus and I.V. Krivosheina,
      Mod. Phys. Lett. A {\bf 21}, 1547 (2006).

\bibitem{Gerda}	J. Jochum (GERDA Collaboration), 
      Prog. Part. Nucl. Phys. {\bf 64}, 261 (2010);
      S. Sch\"onert (GERDA Collaboration), 
      J. Phys. Conf. Ser. {\bf 203}, 012014 (2010).


\bibitem{Rod05} 
      V.A. Rodin, A. Faessler, F. \v Simkovic, and P. Vogel, Phys. Rev. C {\bf 68}, 044302 (2003);
      V.~A.~Rodin, A.~Faessler, F.~\v{S}imkovic and P.~Vogel,  Nucl.\ Phys.  {\bf A766}, 107 (2006);
{\bf A793}, 213(E) (2007). 

\bibitem{anatomy} 
F.~\v Simkovic, A.~Faessler, V.A.~Rodin, P.~Vogel, and  J. Engel, Phys. Rev. C {\bf 77}, 045503 (2008).

\bibitem{suh_UCOM}
  M. Kortelainen, O. Civitarese, J. Suhonen, and J. Toivanen, Phys. Lett. B {\bf 647} 128 (2007);
  M. Kortelainen and J. Suhonen, Phys. Rev. C {\bf 75} 051303 (2007);
  Phys. Rev. C {\bf 76} 024315 (2007).

\bibitem{men09} 
        J. Men\'endez, A. Poves, E. Caurier, and F. Nowacki, Nucl. Phys. A  {\bf 818}, 139 (2009).

\bibitem{Hir08} P.K. Rath, R. Chandra, K. Chaturvedi, P.K. Raina, and J.G. Hirsch,
       Phys. Rev. C {\bf 82},   064310 (2010).

\bibitem{Bar09}
  J.~Barea and F.~Iachello,
   Phys. Rev.  C {\bf 79}, 044301 (2009).

\bibitem{Rodri10}
  T.~R.~Rodriguez and G.~Martinez-Pinedo,
  Phys.\ Rev.\ Lett.\  {\bf 105}, 252503 (2010).

\bibitem{occup}
 F. \v Simkovic, A. Faessler, and P. Vogel, Phys. Rev. C {\bf 77},  015502 (2009).

\bibitem{Delion} D. S. Delion, J. Dukelsky and P. Schuck, Phys. Rev. C{\bf 55}, 2340 (1997);
F. Krmpotic {\it et al.}, Nucl. Phys. {\bf A637}, 295 (1998).

\bibitem{Schuck} J. Dukelsky and P. Schuck, Phys. Lett. {\bf B387}, 233 (1996).

\bibitem{Benes} P. Bene\v{s} and F. \v{S}imkovic,
    AIP Conf. Proc. {\bf 1180}, 21 (2009).
             
\bibitem{Spencer} G. A. Miller and J. E. Spencer, Ann. Phys. {\bf 100}, 562 (1976).

\bibitem{UCOM} H. Feldmeier, T. Neff, R. Roth and J. Schnack, 
   Nucl. Phys. A {\bf  632}, 61 (1998); T. Neff and H. Feldmeier, 
   Nucl. Phys. A {\bf 713}, 311 (2003); R. Roth, T. Neff, 
   H. Hergert, and H. Feldmeier, Nucl. Phys. A {\bf 745}, 3 (2004).

\bibitem{roth} R. Roth, H. Hergert, P. Papakonstantinou, T. Neff and H. Feldmeier,
  Phys. Rev. C {\bf 72}, 034002 (2005). 

\bibitem{src09}  F. \v Simkovic, A. Faessler, H. Muther, V. Rodin, M. Stauf,
  Phys. Rev. C {\bf 79}, 055501 (2009). 

\bibitem{muether} H. M\"uther and A Polls, Phys. Rev. C {\bf 61},
  014304 (1999); Prog. Part. Nucl. Phys. {\bf 45}, 243 (2000).

\bibitem{petcov11} A. Faessler, A. Meroni, S.T. Petcov, F. \v Simkovic, J. Vergados, 
  Phys. Rev. D {\bf 83}, 113003 (2011). 



\bibitem{Vog92} F. Boehm and P. Vogel, Physics of Massive Neutrinos, {\it Cambridge, UK: Univ.Pr. (1992)\\ 249 p.}

\bibitem{SNO+} C.~Kraus and S.~J.~M.~Peeters  [SNO+ Collaboration],
  Prog.\ Part.\ Nucl.\ Phys.\  {\bf 64}, 273 (2010); 
  SNO+ project: \url{http://snoplus.phy.queensu.ca} .

\bibitem{Hir95}  
       J.G.Hirsch, O. Castanos, O. Hess, Nucl.\ Phys.\ A {\bf 582}, 124 (1995).

\bibitem{Esc10}
  A.~Escuderos, A.~Faessler, V.~Rodin and F.~\v{S}imkovic,
 J. Phys. G  {\bf 37}, 125108 (2010).

\bibitem{Sim03}   F.~\v Simkovic, L.~Pacearescu and A.~Faessler,
  Nucl.\ Phys.\  A {\bf 733}, 321 (2004).

\bibitem{Sim04} R. Alvarez-Rodriguez, P. Sarriguren, E. Moya de Guerra, L. Pacearescu, A. Faessler and F. \v Simkovic, Phys. Rev. C  {\bf 70}, 064309 (2004).

\bibitem{Sal09}  M.~S.~Yousef, V.~Rodin, A.~Faessler and F.~\v{S}imkovic,
  Phys.\ Rev.\  C {\bf 79}, 014314 (2009);
  D.~L.~Fang, A.~Faessler, V.~Rodin, M.~S.~Yousef and F.~\v{S}imkovic,
  Phys.\ Rev.\  C {\bf 81}, 037303 (2010).

\bibitem{Fang10}
  D.~L.~Fang, A.~Faessler, V.~Rodin, and F.~\v{S}imkovic,
  Phys.\ Rev.\  C {\bf 82}, 051301(R) (2010).

\bibitem{Fang11}
  D.~L.~Fang, A.~Faessler, V.~Rodin and F.~Simkovic,
  Phys.\ Rev.\  C {\bf 83}, 034320 (2011).

\bibitem{Guess11}
  C.~J.~Guess {\it et al.},
  Phys.\ Rev.\  C {\bf 83}, 064318 (2011).

\bibitem{Bar10}
  A.~S.~Barabash,
  Phys.\ Rev.\  C {\bf 81}, 035501 (2010).

\bibitem{ragha} Chart of nucleus shape and size parameters, \url{http://cdfe.sinp.msu.ru/services/radchart/radmain.html}, and references therein.





\bibitem{frekers}
D.~Frekers,
Prog.\ Part.\ Nucl.\ Phys.\  {\bf 64}, 281 (2010), and contribution in this volume

\bibitem{Rod09} Vadim~Rodin and Amand~Faessler, Phys.\ Rev.\  C {\bf 80}, 041302(R) (2009)

\bibitem{auer72} N.~Auerbach, J.~H\"ufner, A.~K.~Kerman and C.~M.~Shakin,
  Rev.\ Mod.\ Phys.\  {\bf 44}, 48 (1972). 
\bibitem{auer83} N.~Auerbach, Phys.\ Reps. {\bf 98}, 273 (1983).

\bibitem{Rod10}
  V.~Rodin and A.~Faessler,
Prog.\ Part.\ Nucl.\ Phys.\  {\bf 66}, 441 (2011);  arXiv:1012.5176 [nucl-th].

\bibitem{Rod11} 
  V.~Rodin,
  AIP Conf.\ Proc.\  {\bf 1417}, 100 (2011).

\bibitem{tad87}
T.N. Taddeucci {\it et al.},
Nucl. Phys. {\bf A469}, 125 (1987).


\bibitem{Folk75}
F.Folkmann, C.Gaarde 
Nucl.Phys. {\bf A252}, 343 (1975).


\bibitem{Gr07}
 E.~W.~Grewe {\it et al.},
  Phys.\ Rev.\  C {\bf 76}, 054307 (2007).

\bibitem{Dilg73}
W.~Dilg, W.~Schantl, H.~Vonach, and M.~Uhl, Nucl.Phys. {\bf A217}, 269 (1973).



\bibitem{WINTER} R.~Winter, Phys. Rev. {\bf 100}, 142 (1955).

\bibitem{Ver83} J.D. Vergados, Nuc. Phys. B  {\bf 218}, 109 (1983).

\bibitem{BeRuJar83} J.~Bernabeu, A~de~Rujula, and C.~Jarlskog, 
        Phys. Rev. C {\bf 223}, 15 (1983).
 
\bibitem{sujwy}  Z. Sujkowski and S. Wycech, Phys. Rev. C {\bf 70}, 052501 (2004); 
 
\bibitem{lukas}  L. Lukaszuk, Z. Sujkowski and S. Wycech, Eur. Phys. J A {\bf 27}, 63 (2006). 
 
\bibitem{SIM08} F. \v Simkovic, M.I. Krivoruchenko, Phys. Part. Nucl. Lett. {\bf 6}, 298 (2009). 
 
\bibitem{LARKINS} F.B. Larkins, At. Data and Nucl. Data Tables {\bf 20}, 313 (1977).

\bibitem{penning1} G. Douysset \textit{et al.}, Phys. Rev. Lett. {\bf 86}, 4259 (2001).  

\bibitem{penning2} K. Blaum, Phys. Rep. {\bf 425}, 1  (2006). 

\bibitem{penning3} K. Blaum, Yu. N. Novikov, and G. Werth, Contemp. Phys. {\bf 51}, 149 (2010). 

%
\bibitem{kolhinen10} V.S, Kolhinen \textit{et al.}, Phys. Lett B {\bf 684}, 17 (2010).  

\bibitem{mount10} 
  B.J. Mount, M. Redshaw, and E.G. Myers, Phys. Rev. C {\bf 81}, 032501 (2010).

\bibitem{elis4}   M. Goncharov \textit{et al.}, Phys. Rev. C {\bf 84}, 028501 (2011).   

\bibitem{smorra11} C. Smorra \textit{et al.}, arXiv:1201.4942[nucl-ex].

\bibitem{RAKH09} S. Rahaman \textit{et al.}, Phys. Rev. Lett. {\bf 103}, 042501 (2009).  

\bibitem{SCIE09} N.D. Scielzo \textit{et al.}, Phys. Rev. C {\bf 80}, 025501 (2009).  

\bibitem{kolhinen11} V.S, Kolhinen \textit{et al.}, Phys. Lett B {\bf 697}, 116 (2011).  

\bibitem{elis1} S. Eliseev \textit{et al.}, Phys. Rev. Lett. {\bf 106}, 052504 (2011).  

\bibitem{elis3}   S. Eliseev \textit{et al.}, Phys. Rev. C {\bf 84}, 012501 (2011).  

\bibitem{elis2}   S. Eliseev \textit{et al.}, Phys. Rev. C {\bf 83}, 038501 (2011).  

\bibitem{elis5}   S. Eliseev \textit{et al.}, Phys. Rev. Lett. {\bf 107}, 152501 (2011).  

\bibitem{droese11} Ch. Droese \textit{et al.}, submitted to Nucl. Phys. A (2011). 

%
%

\bibitem{ececfang} D.L. Fang, K. Blaum, S. Eliseev, A. Faessler, M.I. Krivoruchenko, 
  V. Rodin, and F. \v Simkovic, Phys. Rev. C {\bf 85}, 035503 (2012).

\bibitem{MAWA1973} J. B. Mann and J.T. Waber, Atomic Data {\bf 5}, 201 (1973).  

%
%
 
\bibitem{barab1}  A.S. Barabash, Ph. Hubert, A. Nachab, and V. Umatov,  
   Nucl. Phys. A {\bf 785}, 371 (2007). 

\bibitem{FPS11} D.~Frekers, P.~Puppe, J.H. Thies, P.~Povinec, F.~{\v S}imkovic, J.~Stani{\v c}ek, and I.~S{\' y}kora,
   Nucl. Phys. A {\bf 860}, 1 (2011).

\bibitem{rukh11} N.I. Rukhadze \textit{et al.}, Nucl. Phys. A {\bf 852}, 197 (2011). 
 
\bibitem{barab2}  A.S. Barabash, Ph. Hubert, A. Nachab, S.I. Konovalov,  
 I.A. Vanyushin, and V.I. Umatov, Nucl. Phys. A {\bf 807}, 269 (2008). 
 
\bibitem{frekdeg} D. Frekers \textit{et al.}, Nucl. Phys. A {\bf 860}, 1 (2011). 

\bibitem{belli12} P. Belli et al., Phys. Rev. C {\bf 85}, 044610 (2012).

\bibitem{belli09} P. Belli \textit{et al.}, Nucl. Phys. A {\bf 842}, 101 (2009).  
 
\end{thebibliography}
\end{document}